\documentclass[hyper]{JHEP3} 

\usepackage{epsfig,
}

\newcommand\fverb{\setbox\fverbbox=\hbox\bgroup\verb}
\newcommand\fverbdo{\egroup\medskip\noindent%
			\fbox{\unhbox\fverbbox}\ }
\newcommand\fverbit{\egroup\item[\fbox{\unhbox\fverbbox}]}
\newbox\fverbbox

\title{Neural Network Parameterizations of Electromagnetic Nucleon Form-Factors}

\author{Krzysztof M. Graczyk\thanks{Supported by: the Ministry of Science and Higher Education project  DWM/57/T2K/2007 as well as  the Polish Ministry of Science Grant  project number:	N N202 368439.}\\
	Institute of Theoretical Physics, Wroc\l aw University,
 pl. M. Borna 9, 50-204 Wroc\l aw, Poland\\
	E-mail: \email{kgraczyk@ift.uni.wroc.pl}}

\author{Piotr P\l onski\\
	Institute of Radioelectronics, Warsaw University of Technology, Nowowiejska 15/19,
00-665 Warsaw, Poland\\
	E-mail: \email{	P.Plonski@stud.elka.pw.edu.pl}}

\author{Robert Sulej\\
	A. Soltan Institute for Nuclear Studies, Hoza 69, 00-681 Warsaw, Poland\\
	E-mail: \email{Robert.Sulej@cern.ch}}

\received{\today} 		
\accepted{\today}		

\abstract{The electromagnetic nucleon form-factors data are studied with artificial feed forward neural networks.
As a result  the unbiased  model-independent form-factor parametrizations are evaluated together with uncertainties.
The Bayesian approach for the neural networks is adapted for $\chi^2$ error-like function and applied to the data analysis. The sequence  of the feed forward neural networks with one hidden layer of units is considered. The given neural network represents a particular form-factor parametrization.  The so-called
\textit{evidence} (the measure of how much the data favor given statistical model)  is  computed with the Bayesian framework and it is used to determine the best form factor parametrization.
}


\begin{document}


\section{Introduction}
\label{section_introduction}

The electromagnetic (EM) form-factors (FF) of the nucleon are the quantities which  embody the
information about the complex electromagnetic structure of the proton and neutron \cite{Close_EM_book}. In  practice, the
form-factors are introduced in order to model (on  effective level)
the electromagnetic hadronic current for elastic $ep\,(n)$ scattering.  In the one photon exchange approximation it has the following form:
\begin{eqnarray}
J_{e\,p(n)}^\mu =\overline{u}(p')\left[ \gamma^\mu F_1^{p(n)}(Q^2)  + \frac{i\sigma^{\mu\nu} q_\nu}{2M_{p(n)}} F_2^{p(n)}(Q^2)\right] u(p),
\end{eqnarray}
where $q_\mu = p'-p$ denotes the four-momentum transfer; $M_{p(n)}$ is the proton (neutron) mass; $p'$  and $p$ are outgoing  and incoming nucleon momenta;  $Q^2 \equiv -q^2$; $F_1^{p(n)}$ is the helicity non-flip Dirac proton (neutron) form-factor, while $F_2^{p(n)}$ denotes the helicity-flip Pauli proton (neutron) form-factor. The form factors are normalized as follows:
\begin{equation}
F_{1}^p(0) = 1, \quad F_{2}^p(0) = \mu_p -1 ,  \quad F_{1}^n(0) = 0, \quad F_{2}^n(0) =\mu_n,
\end{equation}
where $\mu_{p,n}$ is anomalous magnetic moment of the proton, neutron.

The nucleon is the many-body  system of strongly interacting quarks (three valence quarks and any number of quark-antiquark pairs) and gluons. This complex system is  described by the QCD (quantum chromodynamics) in the confinement regime. Study of the EM form-factors gives an opportunity for  testing the models describing the strong interactions. However, computing the EM form-factors from the first principles is an extremely difficult task.  Nevertheless,  some effort has been done with the effective approaches and the lattice QCD.

A good approximation of the FF is performed within the vector meson dominance models  (VMD) \cite{Lomon_VMD,Belushkin:2006qa}. There are interesting results obtained with constituent quark models \cite{quark_model} as well as with other approaches (see for review \cite{Perdrisat:2006hj}). However, the given theoretical description usually  works well only on limited $Q^2$ range. In order to describe the full $Q^2$ domain various approaches must be combined. Hence a proper prediction of the FF in wide $Q^2$ range requires  to use complex phenomenological models which contain    plenty of internal parameters.

On the other hand, the experimental data, which have been collected during the last sixty  years, covers a wide $Q^2$ domain and are
accurate enough to provide reasonable information about the nucleon electromagnetic structure \cite{Miller:2010nz}.
Therefore one can try to represent the nucleon form-factors by the data itself without assuming any model constraints. In this article we follow this philosophy.

Description of the electromagnetic properties of the nucleon is  a problem of great interest of modern particle physics. The knowledge of the nucleon form-factors is also important for  practical  applications. We mention two of them: (i) predicting the cross sections for the quasi-elastic charged current (CC) and elastic neutral current (NC) neutrino scattering off nucleon and nucleus \cite{AlvarezRuso:2009mn};
(ii) investigation of the strange content of the nucleon in
elastic lepton  scattering off nucleons/nuclei \cite{Alberico:2001sd,Beck:2001dz}.

An accurate modeling of the neutrino-nucleus cross sections plays a crucial role in the analysis of the $\nu_\mu\to\nu_\tau$ neutrino oscillation data, collected in the long-baseline experiments. For instance in the experiments like  K2K \cite{Ahn:2006zza} or T2K \cite{T2KLOI} the neutrino energy spectrum is reconstructed from the quasi-elastic-like events. Observing the distortion of the energy spectrum in the far detector gives an indication for neutrino oscillation.

The investigation of  the quasi-elastic CC neutrino-nucleon interactions gives an opportunity to explore the axial structure of the nucleon.
The  weak hadronic  current is formulated  assuming the conserved vector current (CVC) theorem. Then the vector part of the current is expressed in terms of the electromagnetic FF of the proton and neutron, while the axial contribution is described with  two axial form factors: $G_A$ and $G_P$ (pseudoscalar axial form-factor). The hadronic weak current for the CC $\nu n$ quasi-elastic scattering reads \cite{Llewellyn Smith:1971zm}
\begin{equation}
\label{CC_current}
J_{\nu n ,CC}^\mu =\overline{u}(p')\left[ \gamma^\mu F_1^{V}(Q^2)  + \frac{i\sigma^{\mu\nu} q_\nu}{2M} F_2^{V}(Q^2) +
\gamma^\mu\gamma^5 G_A(Q^2) +  \frac{q^\mu}{2M} \gamma^5 G_P(Q^2)\right] u(p),
\end{equation}
where $M=(M_p+M_n)/2$. The isovector Dirac, Pauli form-factors are defined as follows:
\begin{equation}
F_{1,2}^{V}(Q^2) = F_{1,2}^{p}(Q^2)-F_{1,2}^{n}(Q^2).
\end{equation}
If the  partially conserved vector current hypothesis (PCAC) is assumed then the axial  form-factors can be related:
$G_P(Q^2) = 4M^2 G_A(Q^2)/(m_\pi^2 + Q^2)$. The $G_A$   is usually parameterized with dipole functional form:
\begin{equation}
G_A(Q^2) =g_A \left(1 + \frac{Q^2}{M_A^2}\right)^{-2},\quad  g_A = -1.2695 \pm 0.0029.
\end{equation}
$M_A$ denotes the axial mass. Notice that recent studies \cite{Gran:2006jn,
:2007ru} suggest    $M_A$  value  larger by about 20\% with respect to the old measurements
\cite{Bernard:2001rs,Kuzmin:2007kr,Bodek:2007vi}. The impact of the electromagnetic form-factors on the axial mass extraction is small, but it can play a role in the future, when more precise measurements of the neutrino-nucleon cross-sections will be performed.

The precise knowledge of the EM form-factors together with uncertainties is more important for predicting  the NC elastic $\nu N$  reaction cross-section. The structure of the weak NC hadronic current is similar to (\ref{CC_current}) \cite{Alberico:1997vh}, namely:
\begin{equation}
\label{NC_current}
J_{\nu p(n) ,NC}^\mu =\overline{u}(p')\left[ \gamma^\mu F_1^{NC,p(n)}(Q^2)  + \frac{i\sigma^{\mu\nu} q_\nu}{2M_{p(n)}} F_2^{NC,p(n)}(Q^2) +
\gamma^\mu\gamma^5 G_A^{NC,p(n)}(Q^2) \right] u(p),
\end{equation}
where
\begin{eqnarray}
F_{1,2}^{NC,p(n)}(Q^2) &=& \pm \frac{1}{2}F_{1,2}^V(Q^2) - 2 \sin\theta_W F_{1,2}^{p(n)}(Q^2) - \frac{1}{2}F_{1,2}^s(Q^2),\\
G_A^{NC, p(n)}(Q^2) &=& \pm \frac{1}{2} G_A(Q^2) - \frac{1}{2}G_A^s(Q^2),
\end{eqnarray}
$\theta_W$ is the Weinberg angle.   $F_{1,2}^s(Q^2)$ and $G_A^s(Q^2)$ describe the strange content of the nucleon.
We see that the investigation of the elastic NC neutrino-nucleon scattering gives the opportunity to explore the nucleon strangeness
\cite{Alberico:1997vh,Kim:2008zzc} (mainly the axial strange part). The strangeness of the nucleon is also
investigated in the elastic $ep$ scattering \cite{Beck:2001dz,strangness}.  The extraction of this contribution
is sensitive to  the accuracy of the EM form-factors. Therefore it is necessary to use the well determined FF parametrization together with the uncertainties.

There are many  different phenomenological parametrizations of the EM
form-factors  \cite{Belushkin:2006qa,Bosted:1994tm,Brash:2001qq,Budd:2003wb,Arrington:2003df,Kelly:2004hm,Arrington:2006hm,Bodek:2007ym,Galster:1971kv,Krutov:2002tp}. Some of these are based on the theoretical models, but mostly in practical applications simple functional parametrizations fitted to the data are applied \cite{Alberico:2008sz}.
The functional form  is chosen to satisfy some general  properties (proper behavior at $Q^2\to0$ and
$Q^2\to\infty$, scaling behavior). However, a particular choice of the parametrization
determines the final fit and affects also the uncertainty. The form-factors parameterized by the large number of degrees of freedom
have a tendency to describe the data too accurately, and the generality of the fit is lost. On the other hand,  the model with a small
number of the parameters may describe the data imprecisely. Moreover  the complexity of the fit  has an impact on its
uncertainties.

Searching for the proper parametrization, which describes the data well enough without losing the  generality of the fit is just solving the problem, known in statistics as \textit{bias-variance trade-off} \cite{Bishop_book,Geman92}.  Usually the most reasonable solution is chosen with a use of \textit{common sense}, i.e. the fit which leads to the low enough $\chi^2_{min}$ value is accepted, and  more complex models are not  considered. The task of this paper is to evaluate a model independent FF parametrizations, which will not be affected by the problems described above i.e. the common sense will be replaced by the objective Bayesian procedure.

One of the possible fitting techniques is to apply artificial neural networks (ANN).  The ANN has already been used in the high energy physics for decades \cite{NN_hep_old} and it has been shown to be a powerful tool in the field. The pattern recognition tasks like particle or interaction identification are efficiently addressed with the ANN based methods also in present experiments \cite{NN_classify,NN_compass}. The ANN are also applied to the function approximation and parameter estimation problems \cite{Forte:2002fg,NN_esimator}.

The ANN techniques have already been applied by  NNPDF collaboration \cite{NNPDF} to represent the nucleon and
deuteron EM structure functions \cite{Forte:2002fg,Del Debbio:2004qj,DelDebbio:2007ee,Ball:2008by,Ball:2009qv,Ball:2010de}.
The method is based on the large collection of networks \cite{Ball:2008by} of the same architecture  prepared on the artificial data sets generated from original experimental measurements. Obtained fits are claimed to be unbiased  due to networks being intentionally oversized -- the number of free parameters, the network \textit{weights}, is  larger than required to solve the problem. To avoid potential over-fitting (representing the statistical fluctuations of the experimental data), that may arise under  these conditions, the optimization of the network weights (so-called \textit{training}) is stopped before reaching the minimum of the figure of merit (\textit{error function}) calculated on training data. Stopping condition is based on the cross-validation technique, where the portion of available data is excluded from training. Such created subset is then used to calculate the test error function which starts to increase when the network becomes fitted to training data more than to the testing data. This observation is used to break the training. The best fit values and the uncertainties given by the NNPDF are computed by taking the average and standard deviation respectively, over the set of solutions obtained from the whole collection of the networks.

In the case of the present analysis  the number of experimental points varies from 26 to 57, and we do not generate
the Monte Carlo data. Therefore the cross-validation technique
is unsuitable because constructing the testing data set can significantly restrict the information about the underlying data model
used in training. Additionally our intention is to compare statistical models which are represented by the networks of various
architectures and among them choose the most appropriate parametrization. It motivated us to consider another idea for finding the best fit and the choice  of the neural network architecture. We apply Bayesian framework (BF)  for the ANN.   It is a different philosophy of building the statistical model  than the NNPDF approach. However, both techniques are complementary and face with the same \textit{bias-variance trade-off}. A pedagogical description of the main ingredients of both methodologies can be found in  Ch.~Bishop's book \cite{Bishop_book} (chapters 9 and 10 respectively).

In the BF approach the sequence of neural networks characterized by different number of hidden units is considered.
A given network of a particular size has its specific ability to adjust to training data i.e. small networks give smooth approximation,
large networks can over-fit the data. One can think that the network of a particular architecture represents
the particular statistical model. With the help of the Bayesian technique we compare the models  and choose
the most appropriate one. This method has been developed for the ANN \cite{Bishop_book,MacKay92a,MacKay92b,MacKay94} in nineties of
last century. We adapted this approach for the purpose of $\chi^2$ minimization. In practice, the so-called \textit{evidence} is computed for every network type in order to select the most appropriate parametrization for given data set.  The evidence is a probabilistic measure which indicates the best solution.

The network of particular architecture has weights that need to be optimized i.e. the
global minimum of the error function is searched for. In order to get the solution  we consider various gradient algorithms.
However, the  training done with these algorithms   can  stick in local minimum. Therefore for a given network architecture
the sample of networks with randomized initial weights is trained to find a single configuration at the global minimum
(this procedure is described in Sec. \ref{section_training_networks}).
The error function is modified with so-called \textit{regularization} term to improve
generalization ability (to control the over-fitting); the extent of regularization
is controlled in the statistically optimal way, also as a part of the Bayesian algorithm.

The main results of our studies are unbiased proton and neutron FF parametrizations, available in the numerical
form at \cite{sulej_network} as well as in the analytical ones  (see Appendix A).
The proposed statistical method also allows to compute  the form-factor uncertainties (from the covariance matrix). One of the strengths of this methodology is its ability of studying the deviations of the form-factors
from the dipole form.

Eventually, let us mention that the previous (non-neural) form-factor data analysis (with ah-hoc parametrizations)  have been done in  the non-Bayesian spirit i.e. authors do not compare  the possible FF parametrizations in order to choose the most suitable. Usually the one particular functional form was discussed and analyzed with the  $\chi^2$ framework.

The paper is organized as follows. In Sec. \ref{section_feed_forward} the feed forward neural networks are shortly reviewed.
Sec. \ref{section_bayes} describes the Bayesian approach to neural networks. The last section contains the numerical results
and discussion. We supplement the article with the appendix, which presents the fits in the analytical form.

\section{Feed Forward Neural Networks}
\label{section_feed_forward}

\subsection{Multi-Layer Perceptron}
\label{section_feed_forward_mlp}
\DOUBLEFIGURE[pos]{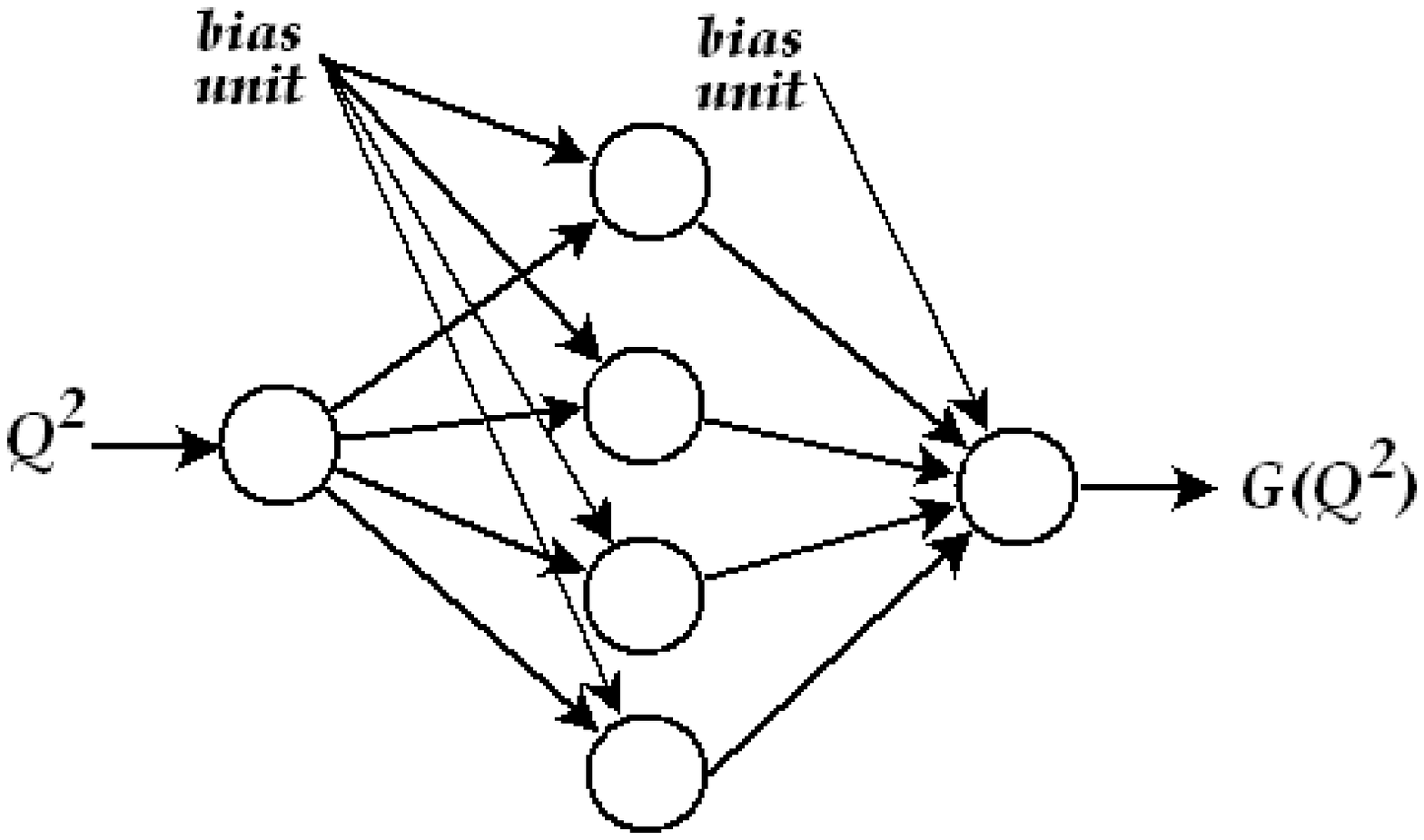,width=7cm, height=5cm}{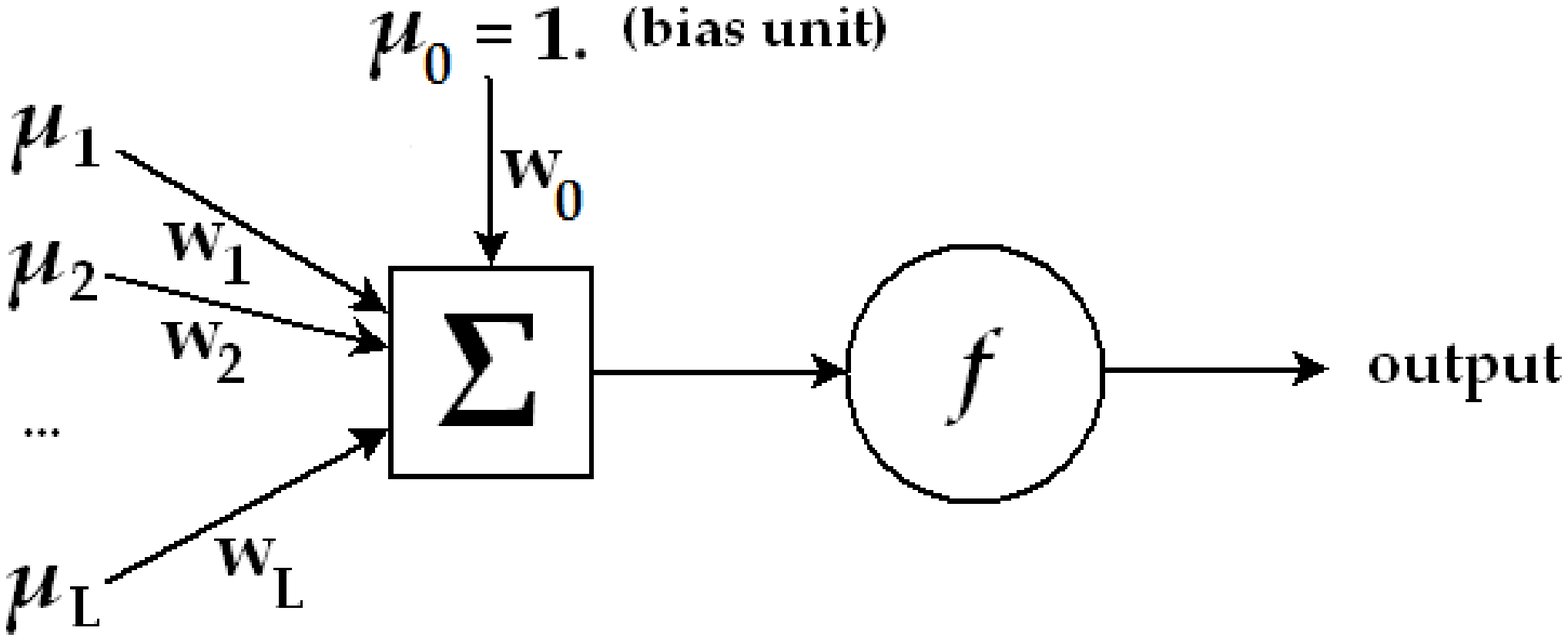,width=6cm, height=5cm}{The feed forward neural network (of type 1-4-1) with one hidden layer, one input and
output unit and 4 hidden units, representing the form-factor $G(Q^2)$. \label{siec_fig}}{Single neuron.  \label{neuron_fig}}
We consider the feed-forward neural network  in the so-called multi-layer perceptron (MLP) configuration.
The network structure (shown in Fig.~\ref{siec_fig}) contains: the input layer, the layer of $M$ hidden neurons
and a single neuron in the output layer. We will say  that the network  of type 1-M-1  is considered.
Each neuron  (see Fig. \ref{neuron_fig}) calculates the output value as an activation function $f_{act}$ of
the weighted sum of its inputs:
\begin{equation}
f_{act}\left(\sum_i w_i \mu_i\right),
\end{equation}
where $w_i$ denotes the weight parameter, while $\mu_i$ represents the output value of the unit from previous layer.
Neurons in the hidden layer are usually non-linear, with the sigmoid or hyperbolic tangent functions denoted  as $f_{act}$;
in this analysis the output neuron is linear function. In  general, the ANN gives a map ($\vec{y}$) of the input into the output
vector spaces. The overall network response  is then a deterministic function of the input
variable (vector $\vec{in}$), and the weight parameters:
\begin{equation}
\vec{y}(\vec{in},\vec{w}): \; \mathcal{R}^{D_{input}} \to \mathcal{R}^{D_{output}}.
\end{equation}
In our analysis  the ANN is  expected to approximate the given form-factor $G$ depending on  the input variable $Q^2$:
\begin{equation}
\label{nn_output}
y(Q^2, \vec{w}) = G(Q^2).
\end{equation}

Let $\mathcal{D}$ denotes the training data set of $N$ points:
\begin{equation}
\mathcal{D} = \{(x_1 , t_1, \Delta t_1 ), ...,  (x_i , t_i, \Delta t_i ), ..., (x_N , t_N, \Delta t_N )  \},
\end{equation}
where $t_i$ is the measured value of the nucleon form-factor at the point $x_i=Q^2_i$, while the $\Delta t_i$ denotes the total experimental error. The network training goal is to find $\vec{w}$ that minimizes an error function defined here as:
\begin{equation}
\label{total_error_function}
S(\vec{w},\mathcal{D}) = \chi^2(\vec{w},\mathcal{D}) + \alpha E_w(\vec{w}).
\end{equation}
$\chi^2$ term is the error on data:
\begin{equation}
\label{chi2_definition}
\chi^2(\vec{w},\mathcal{D}) = \sum_{i=1}^N \left(\frac{y(x_i,\vec{w}) - t_i}{\Delta t_i} \right)^2.
\end{equation}
$\alpha$ parameter is the factor for the regularization term $E_w$. In this work we apply the weight decay formula
\cite{NN_weight_decay}:
\begin{equation}
\label{regularyzator_definition} E_w(\vec{w}) = \frac{1}{2}\sum_{i=1}^W w_i^2,
\end{equation}
where $W$ denotes the total number of weights in the network (including bias weights).


In general, the output of the MLP with $M$ hidden neurons and the linear output neuron can be written in the form:
\begin{equation}
y(\mu_0,...,\mu_L) = \sum_{m=0}^{M}\left[ w_m^{out} f_{act}\left(\sum_{l=0}^L w_l^{hid_m} \mu_l\right) \right].
\end{equation}
In this paper we consider the neural networks with  $(L=1)$: one input unit $\mu_1 = Q^2$, and one bias unit $\mu_0=1$ in the first layer. The bias of the output neuron in the above formula is considered as the hidden neuron with the constant output, $f_{act}=1$. Such representation
closely corresponds to the Kolmogorov function superposition theorem \cite{Kolmogorov}. Basing on this relation it
was shown \cite{NN_approx1, NN_approx2} that the MLP can approximate any continuous function of its inputs,
to the extent that depends on the number of the hidden neurons. However, in the practical problem we are faced,
the desired function is not known and only the limited number of experimental points is available instead. It
leads to the mentioned earlier \textit{bias-variance} problem. The output of the oversized network tends to approach
closely to the training data points if weights are not constrained during the training. Usually this means
 that statistical fluctuations are captured. The weight regularizing term (Eq. \ref{regularyzator_definition})
 penalizes the large weight values and smooths out the network output, but on the other hand, applying the regularization with overestimated
 value of the factor $\alpha$ leads to the fit which does not reproduce significant features of the training data.
 The effect of applying regularization is illustrated in Figs.~\ref{Fig_regularyzator} and \ref{Fig_regularyzator2},
 where the relatively large network was trained with various values of the factor
 $\alpha$. Similarly, the network with the low number of the hidden neurons may be not capable to represent the desired function.
 Sec. \ref{section_bayes} presents the statistical approach to  determine the network size appropriate to the given data set and to predict
 the optimal value of $\alpha$.
\DOUBLEFIGURE[pos]{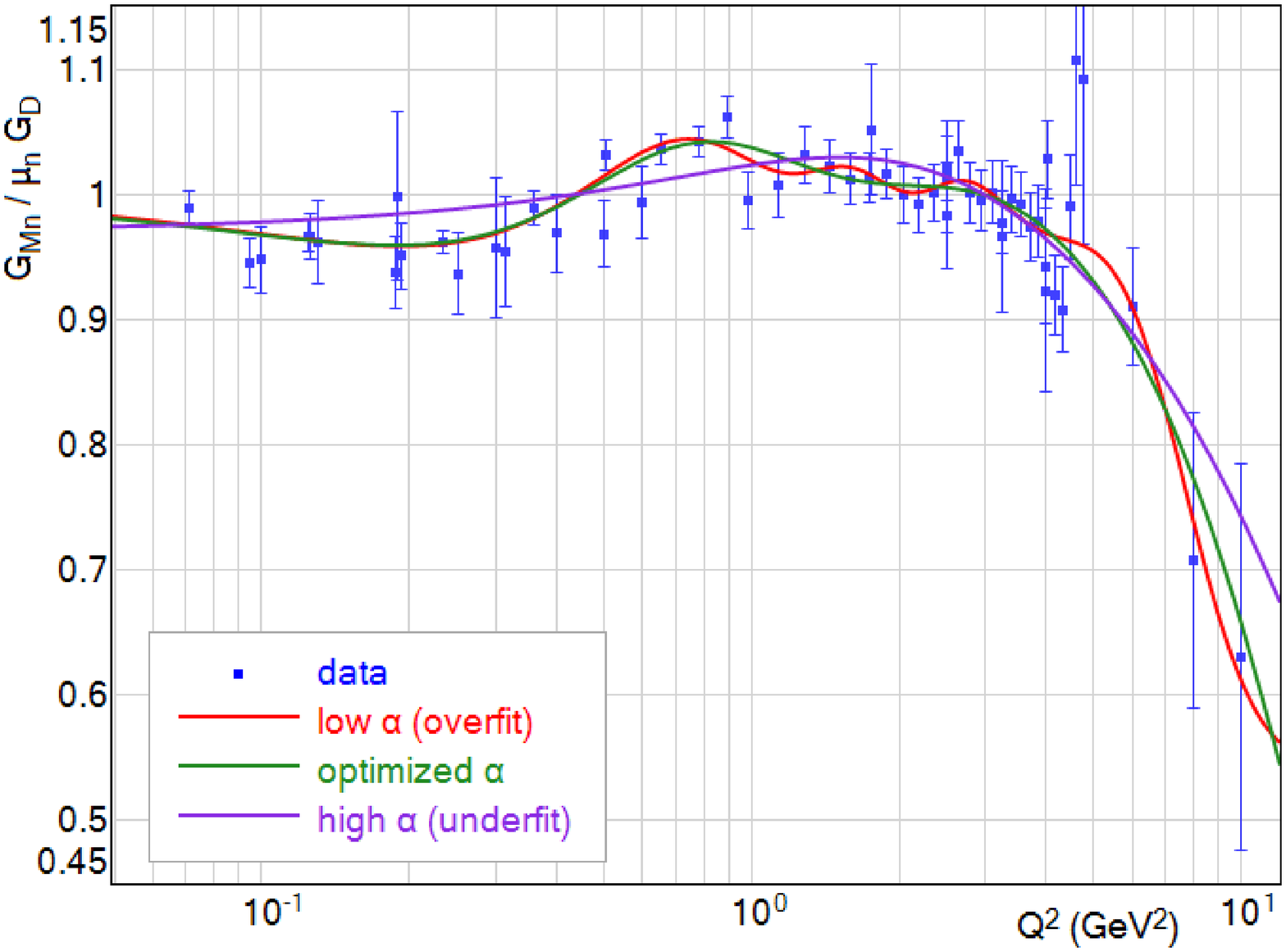,width=0.5\textwidth }{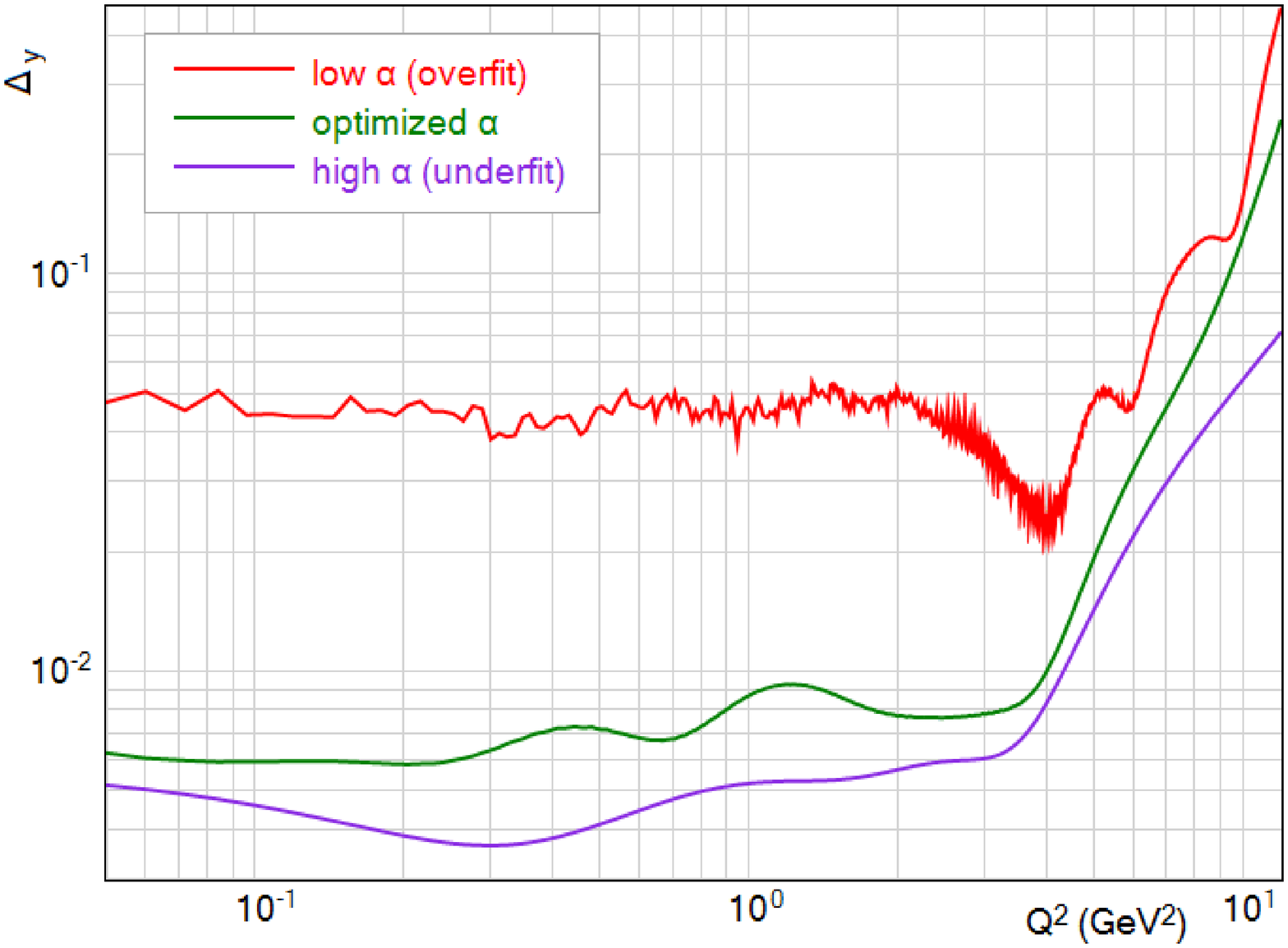,width=0.5\textwidth}
{
Fits of the $G_{Mn}/\mu_n G_D$ data parametrized with the network of large size. The results were obtained with: fixed, underestimated value of $\alpha$ (red line); fixed, overestimated value of $\alpha$ (violet line); online optimized value of $\alpha$ (green line). \label{Fig_regularyzator} }
{The $G_{Mn}/\mu_n G_D$  uncertainties (of the fits presented in  Fig. 3) computed with (3.16). \label{Fig_regularyzator2}}

\subsection{Training of Network}

\label{section_training_networks}

It has been already mentioned that the training of the network is the process of establishing $\vec{w}$ which minimizes the error function (\ref{total_error_function}). We denote the minimal error by $S(\vec{w}_{MP},\mathcal{D})$  (the notation will become clear latter).

The first algorithm for the MLP weights optimization, the \textit{back-prop}, was proposed by D. E. Rumelhart \textit{et al.} in \cite{mlp}.
Currently there is a wide range of gradient descent and stochastic algorithms available for the network training. We use mainly the \textit{Levenberg-Marquardt} algorithm \cite{NN_Levenberg,NN_Marquardt}, since it converges efficiently and does not require precise parameters tuning. However, we trained the networks also with \textit{quick-prop} \cite{NN_quickprop}, and \textit{rprop} \cite{RPROP} algorithms. The obtained results were very similar.

The algorithms we use, as all gradient based optimization patterns, may suffer from local minima.
Therefore for given network type 1-M-1 we consider  a large sample of networks with different (randomized) initial weights.
We use a limited range of initial weight values according to the properties of the neuron activation
function\footnote{High weight values make the sigmoid activation function very steep.
Then the neuron input values have a very narrow range, where the neuron output is not saturated -- this would efficiently
block the training, where the output derivative is used extensively.
Hence we restrict the initial weight range to $|w_{initial}|=f^{sat}_{act}/(L\overline{\mu})$,
where $f^{sat}_{act}$ is the value for which activation function saturates,
$L$ is the number of neuron inputs, $\overline{\mu}$ is the mean neuron input value.}.

After the training of the sample of networks of the same type the distribution of the total error value
$S(\vec{w}_{MP},\mathcal{D})$ is obtained (see Figs. \ref{Fig_GEP_chi2} and \ref{Fig_GMP_chi2}).
Notice that the distribution sharply starts at particular $S_{cut}$ value. Such clear cut on the error value gives us
an indication that the global minimum is well approximated. The number of networks in the sample required to determine
the clear $S_{cut}$ value depends on the complexity of the data. The typical number we obtained were as follows: 150 ($G_{En}$ data), 250 ($G_{Mp}$ data), 700 ($G_{Mn}$ data), and 1300 ($G_{Ep}$ data).

The Bayesian framework  allows to choose from the sample \textit{the best model}. It is the solution characterized by the highest evidence (as it is described in Sec. \ref{section_bayes}). In practice, if the total error is too big then the evidence is too low and the given network can be discarded from further analysis. Hence to simplify the numerical procedure we take into consideration ten fits (neural networks) with the lowest total error values. They are also characterized by the low $\chi^2$ value, namely  $\chi^2(\vec{w}_{MP},\mathcal{D})/(N-W)  <  1$; $N-W$ is the number of degrees of freedom. Among them the one with the maximal evidence is selected for further comparison with the network of other types. It was interesting to observe that the fit parametrizations given by the average over the fits selected by lowest error value were found to be very similar to those indicated by the highest evidence in each sample. This observation confirms that all solutions we select from the sample are localized in close neighborhood of the global minimum and are very similar to the one indicated by  the highest evidence.

\DOUBLEFIGURE[pos]{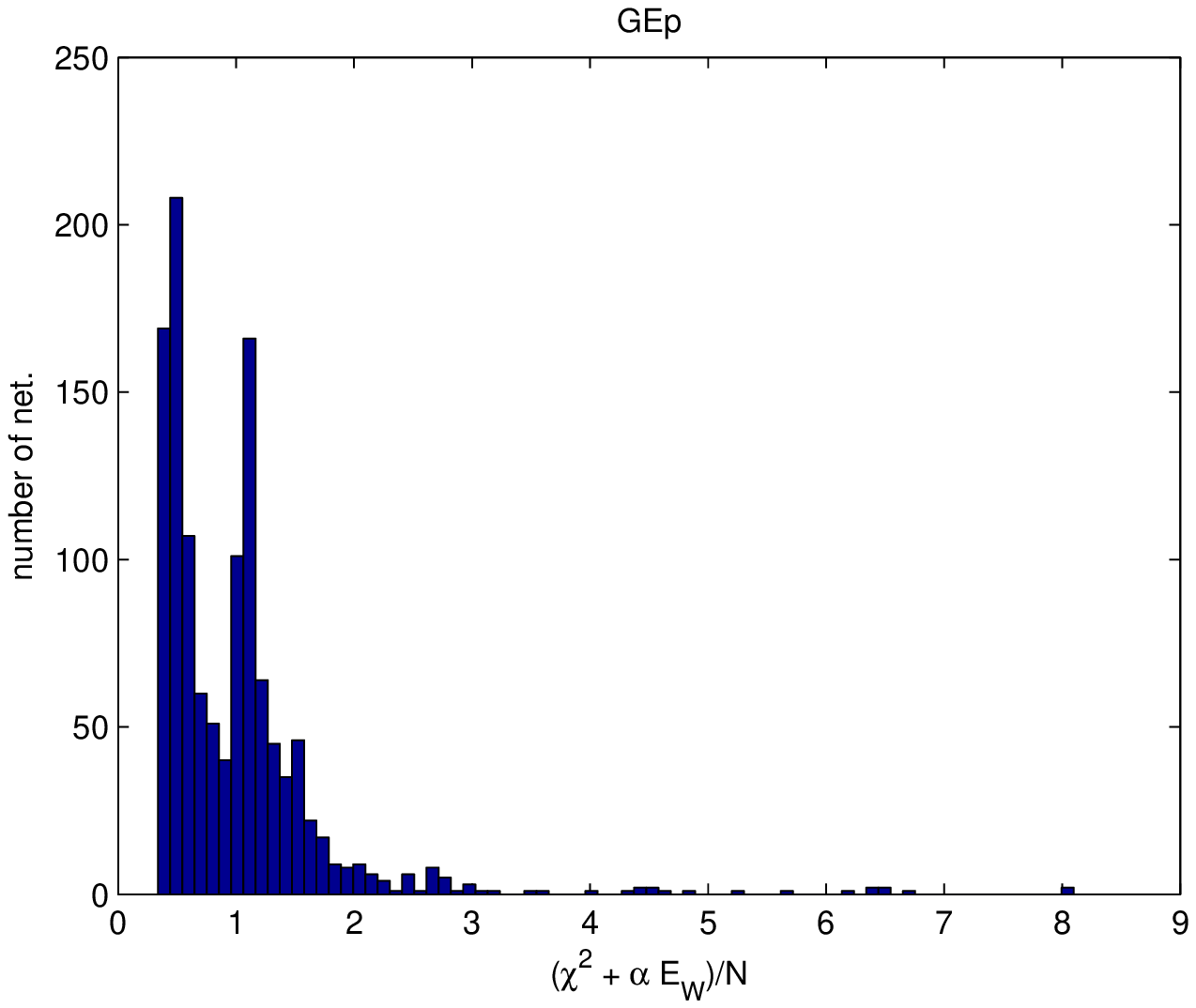,width=0.5\textwidth}{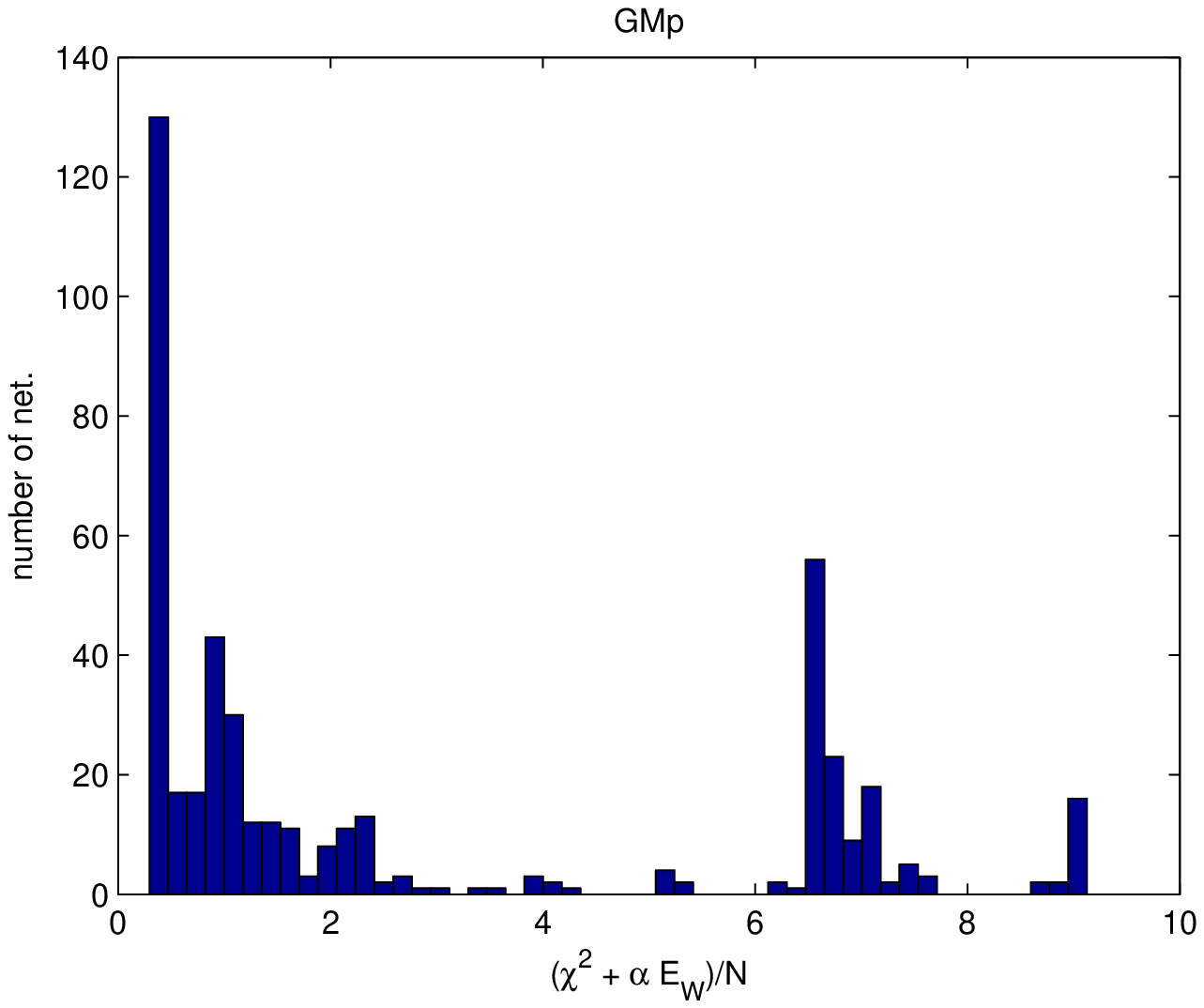,width=0.5\textwidth}{ $S(\vec{w}_{MP},\mathcal{D})/N$ distribution obtained for the network sample trained with the $G_{Ep}/G_D$ data. The 1-3-1 network type was applied. \label{Fig_GEP_chi2}}{
 $S(\vec{w}_{MP},\mathcal{D})/N$ distribution obtained for the network sample trained with the $G_{Mp}/\mu_p G_D$ data. The 1-3-1 network type was applied. \label{Fig_GMP_chi2}}

\section{Bayesian Approach to Neural Networks}
\label{section_bayes}

\EPSFIGURE[pos]{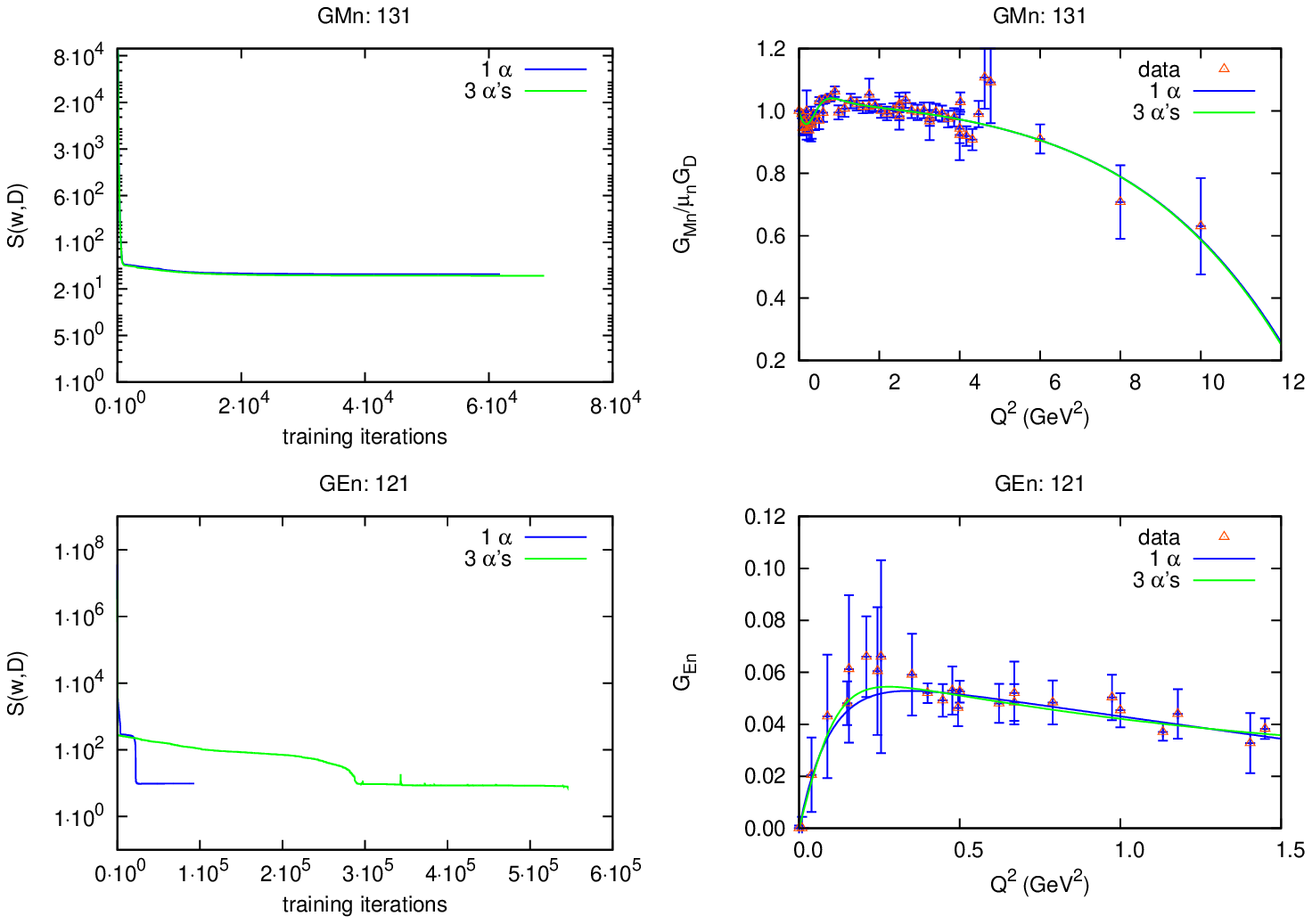}{Left panels: $S(\vec{w},\mathcal{D})$ dependence on the iteration step. Right panels: the best fits obtained for $G_{Mn}/\mu_n G_D$ and $G_{En}$ data. The results obtained with (\ref{prior_3_alpha}) prior are denoted by green lines, while the results computed for the (\ref{prior_alpha}) prior function are plotted with blue lines.  For the magnetic neutron data the network of 1-3-1 type was trained. The electric neutron data was analyzed with 1-2-1 network type.
\label{fig_alpha_regula}}

\EPSFIGURE[pos]{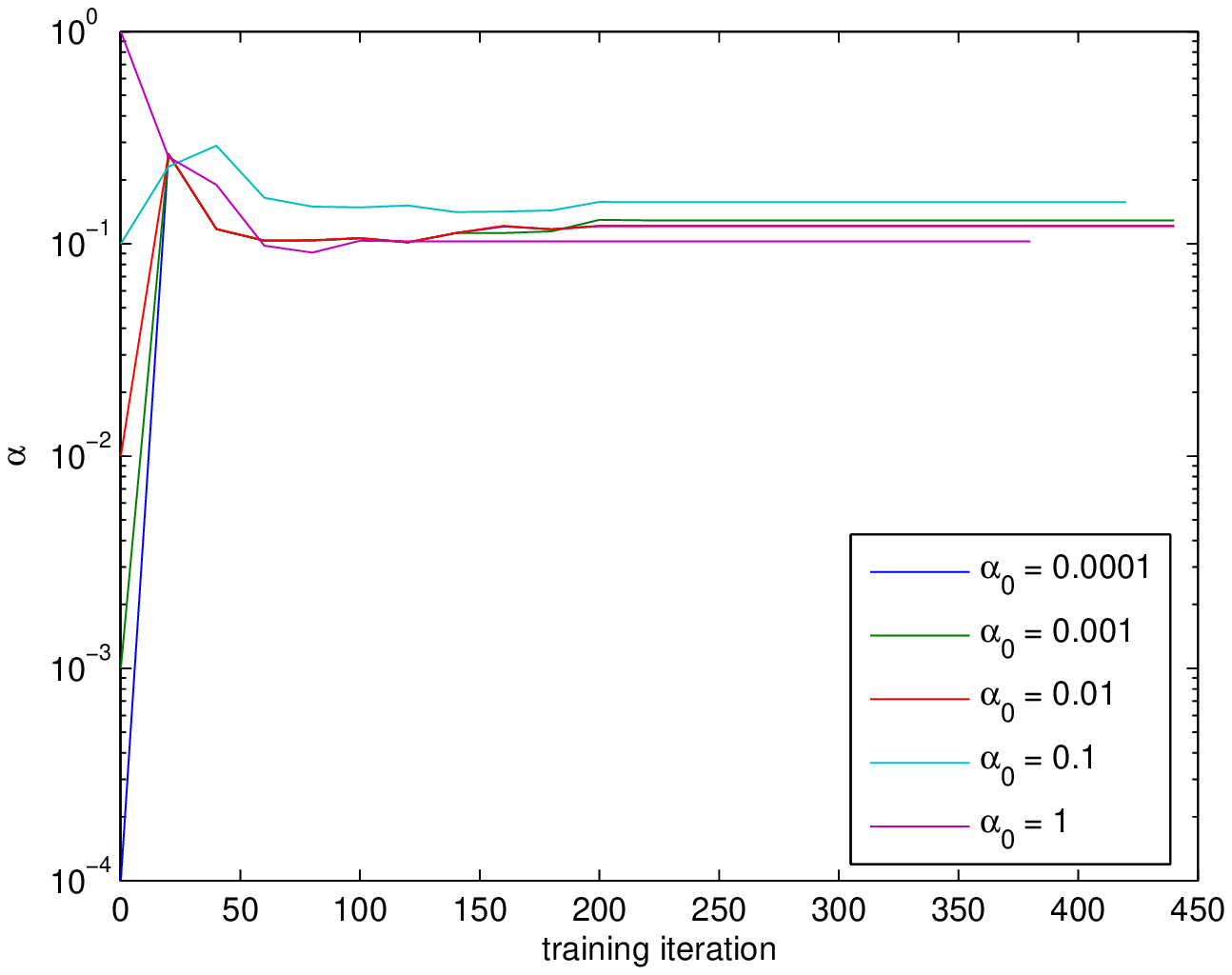,width=0.6\textwidth}{Dependence of iteration of $\alpha$ parameter on the initial $\alpha_0$ value. The results were obtained for the
1-2-1 network type trained with $G_{En}$ data. \label{Fig_alpha_iter}}

The Bayesian framework (BF) for the model comparison \cite{MacKay92a,MacKay92b,MacKay94,Bishop_book,Thodberg}
is taken into consideration. We adapt this framework for  $\chi^2$ minimization purpose. The data is analyzed with the set of various neural networks types $\mathcal{A}_M$: 1-M-1.
Given neural network of architecture $\mathcal{A}_i$ corresponds to  a particular statistical model (hypothesis) describing data.
The BF  allows to:
\begin{itemize}
\item quantitatively classify the hypothesis;
\item choose objectively the best model (neural network) for representing a given data set;
\item establish objectively the weight decay parameter $\alpha$ (see Eq. \ref{regularyzator_definition});
\item compute the uncertainty for the neural network response (output), and uncertainties for other network parameters.
\end{itemize}
The approach in  natural way embodies the so-called Occam's  razor criterium which penalizes more complex models and prefers simpler
solutions.
\EPSFIGURE[pos]{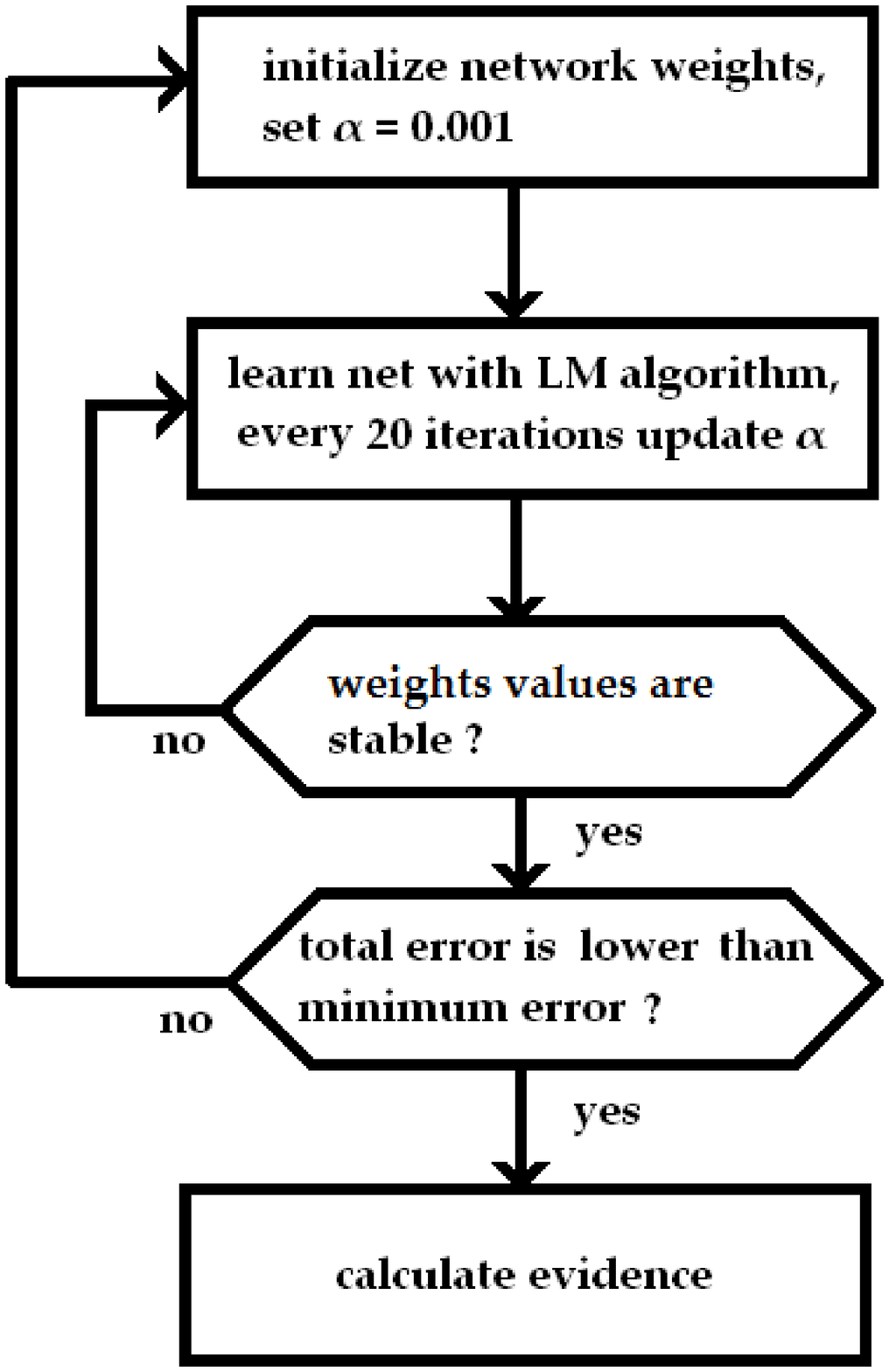,width=0.6\textwidth}{Learning schema. \label{Fig_schemat_blokowy}}

\subsection{Bayesian Algorithm}

\DOUBLEFIGURE[pos]{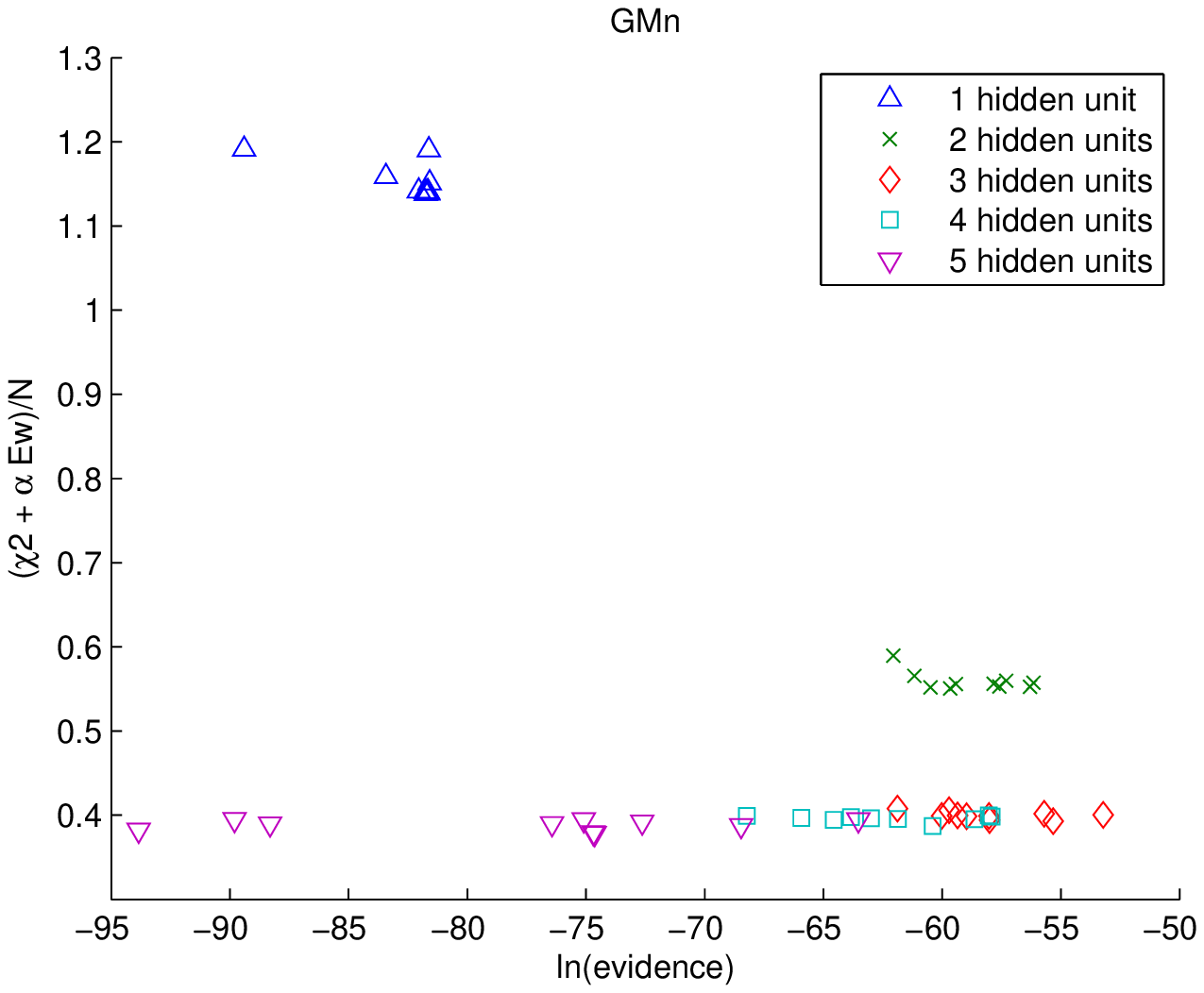,width=0.5\textwidth}{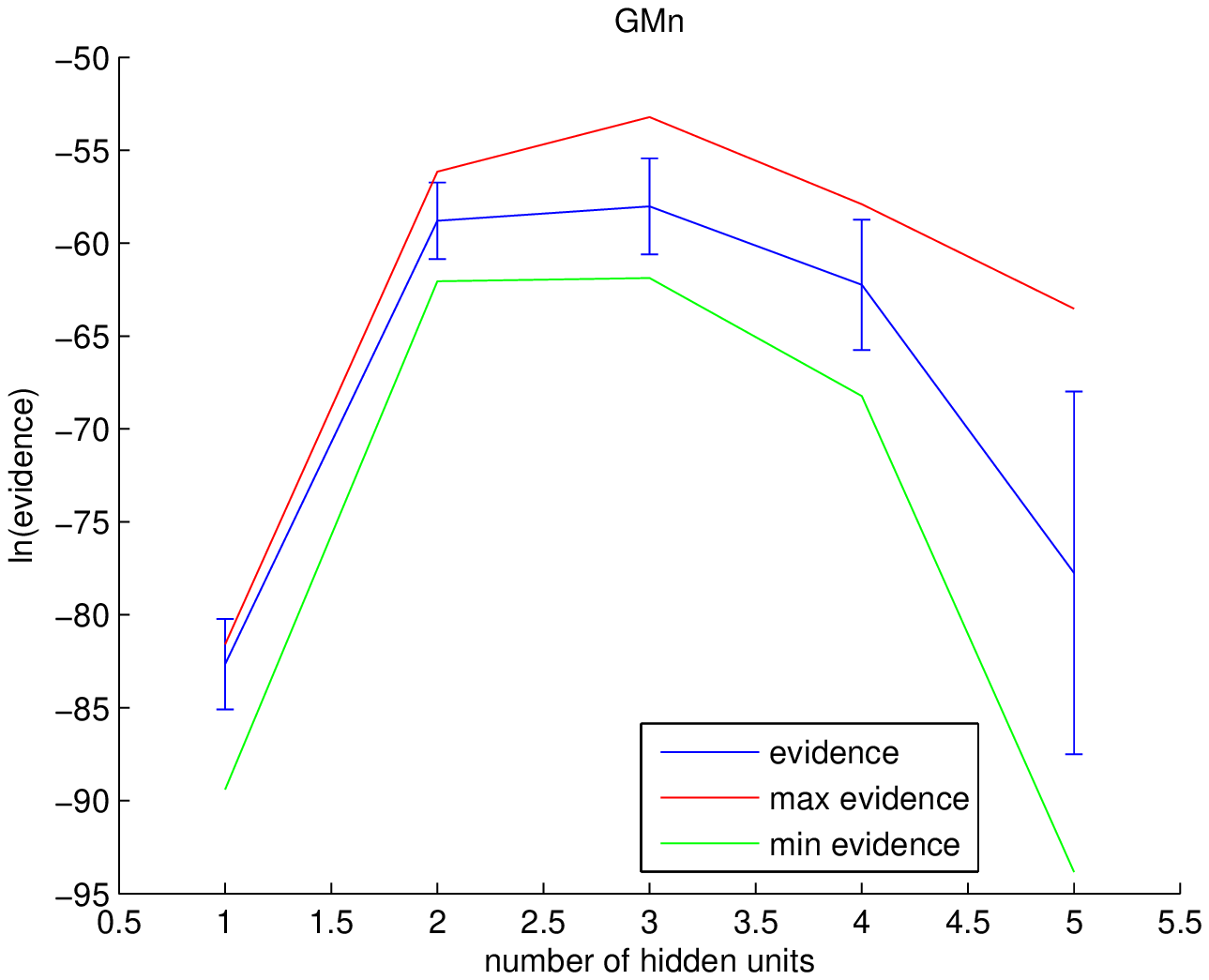,width=0.5\textwidth}
{ The total error, $S(\vec{w}_{MP})$, as a function of $\ln \mathcal{P}\left(\mathcal{D}\right|\left. M \right)$ (ln evidence). The evidence is computed for networks trained with $G_{Mn}/\mu_n G_D$ data. The results obtained for  networks with $M=1-5$ hidden units are shown. Single point represents the fit obtained for given starting weight configuration and particular network type.
\label{Fig_evidence_GMn}
}
{
The dependence of
$\ln \mathcal{P}\left(\mathcal{D}\right|\left. M \right)$ on the number of hidden units. The evidence is computed for networks trained with $G_{Mn}/\mu_n G_D$ data. The maximal  and minimal values of $\ln \mathcal{P}\left(\mathcal{D}\right|\left. M \right)$ (for given network type) are plotted with the red and green lines respectively.  The mean of $\ln \mathcal{P}\left(\mathcal{D}\right|\left. M \right)$ over all acceptable solutions  is represented by the blue line.
\label{Fig_evidence_GMn2}
}
\DOUBLEFIGURE[pos]{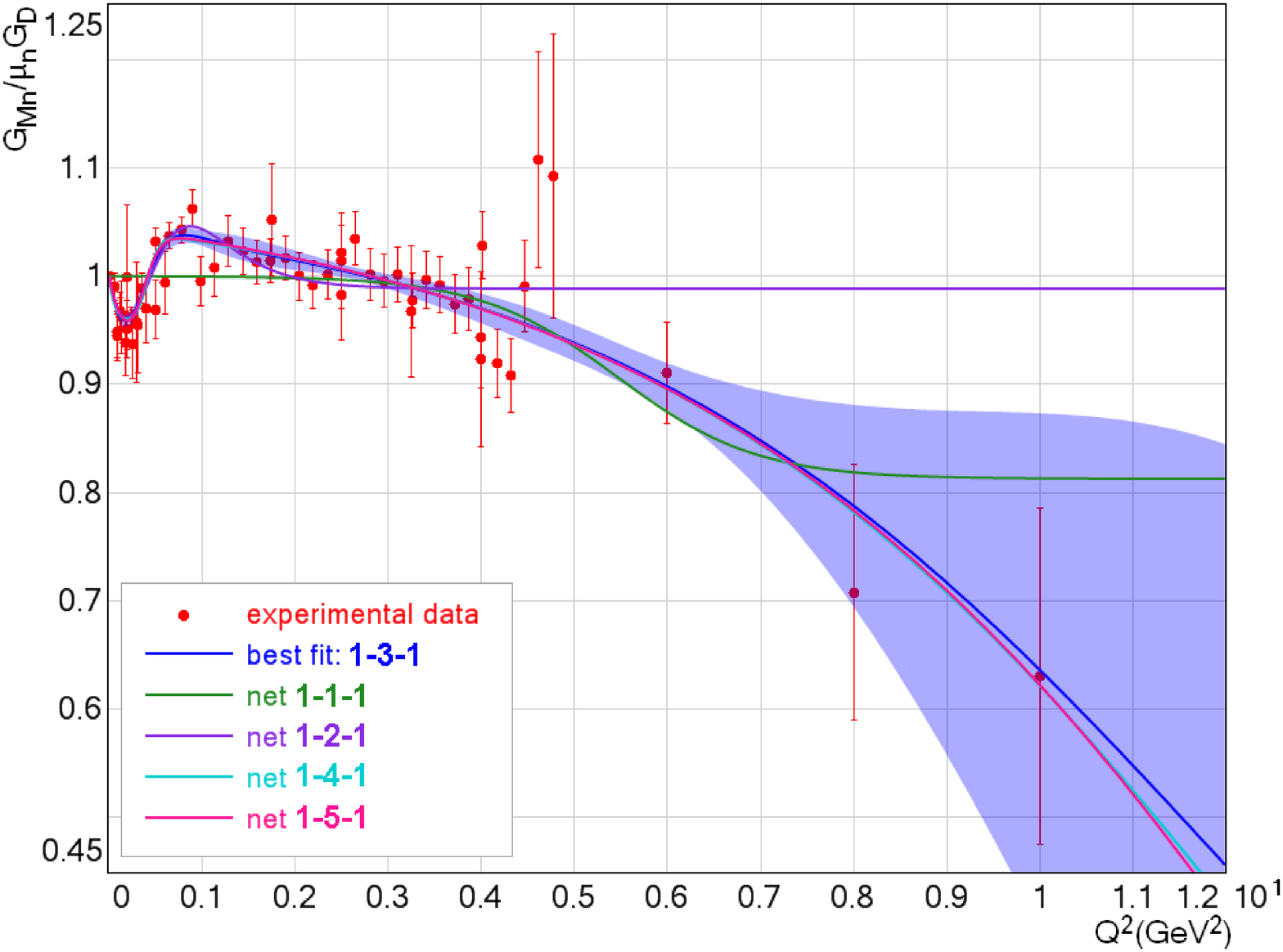,width=0.5\textwidth}{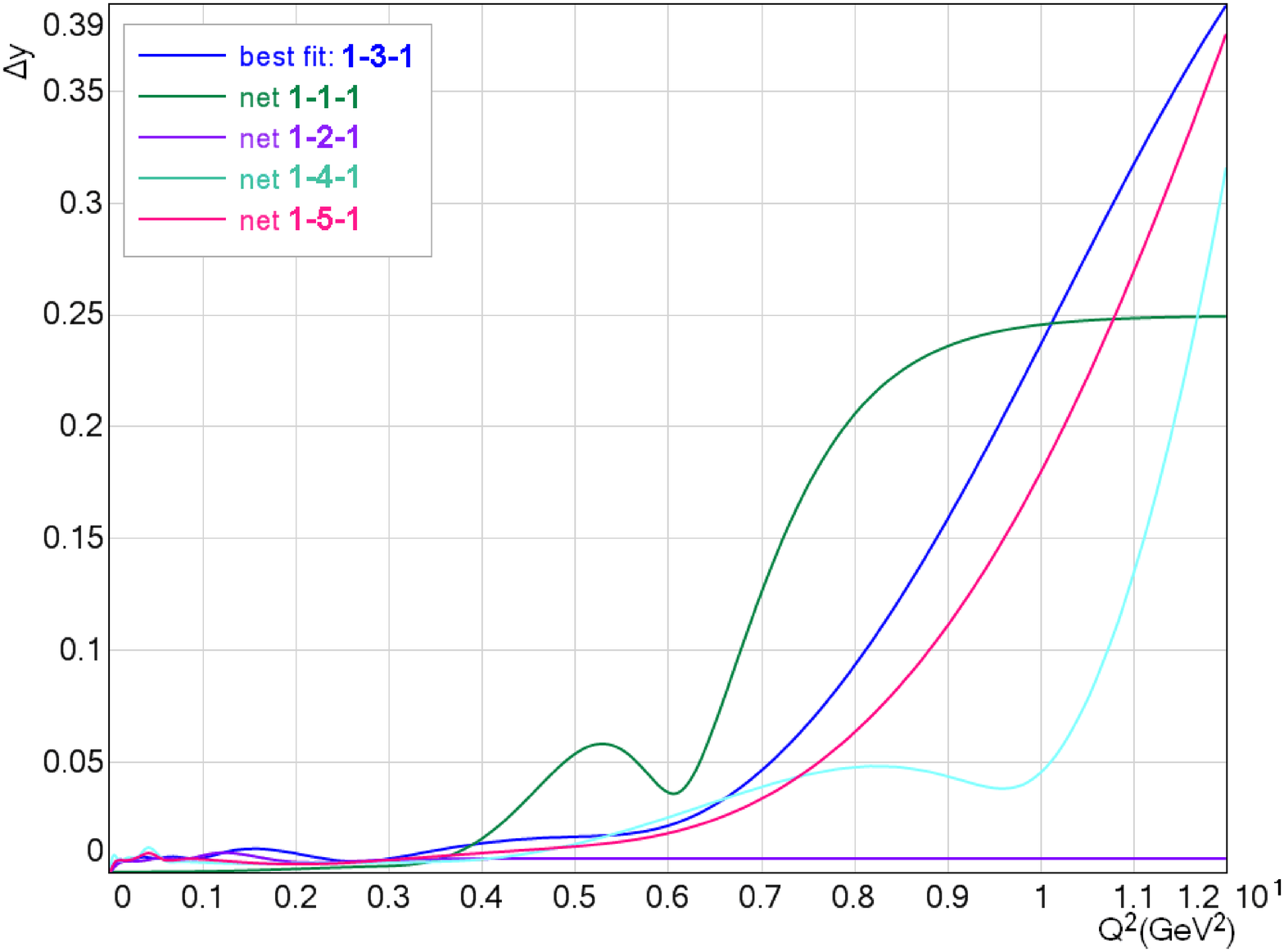,width=0.5\textwidth}{Fits of the $G_{Mn}/\mu_n G_D$ data
parametrized with networks of 1-1-1 (green line), 1-2-1 (violet line), 1-3-1 (blue line), 1-4-1 (cyan line) and 1-5-1 (magenta line) types. The best fit (shown with 1$\sigma$ uncertainty), which was indicated by the maximal evidence,  is given by 1-3-1 network. The blue area denotes fit uncertainty computed with (3.16).  The experimental data is the same as the one discussed in Ref. \cite{Alberico:2008sz}. \label{Fig_GMn}}{The fit uncertainty computed (with Eq. 3.16) for the parametrizations shown  in Fig. 12.   \label{Fig_GMn2}}

At the beginning of the fitting procedure  every neural network architecture $\mathcal{A}_M$
is classified by the prior probability $\mathcal{P}(\mathcal{A}_M)$. After the training of the network with the data $\mathcal{D}$, the posterior probability is  evaluated
$\mathcal{P}\left(\mathcal{A}_M\right|\left.  \mathcal{D}\right)$ i.e. a probability of the model $\mathcal{A}_M$ given data $\mathcal{D}$. It classifies quantitatively considered hypothesis.

On the other hand applying the Bayes' theorem allows to express the posterior probability in the following way:
\begin{equation}
\mathcal{P}\left(\mathcal{A}_M\right|\left.  \mathcal{D}\right) =
\frac{\mathcal{P}\left(\mathcal{D}\right|\left. \mathcal{A}_M \right) \mathcal{P}(\mathcal{A}_M)}{\mathcal{P}(\mathcal{D})},
\end{equation}
where:
\begin{equation}
\label{evidence_def}
\mathcal{P}\left(\mathcal{D}\right|\left. \mathcal{A}_M \right)
\end{equation}
is called evidence \cite{MacKay92a} (probability of the data $\mathcal{D}$ given $\mathcal{A}_M$).

There is no reason to prefer some particular model before starting  data analysis, hence:
\begin{equation}
\mathcal{P}(\mathcal{A}_1) = \mathcal{P}(\mathcal{A}_2) =...= \mathcal{P}(\mathcal{A}_M) =...
\end{equation}
Then if one neglects the normalization factor $\mathcal{P}(\mathcal{D})$
the evidence (\ref{evidence_def}) is the probability distribution which quantitatively  classifies  hypothesis.
\DOUBLEFIGURE[pos]{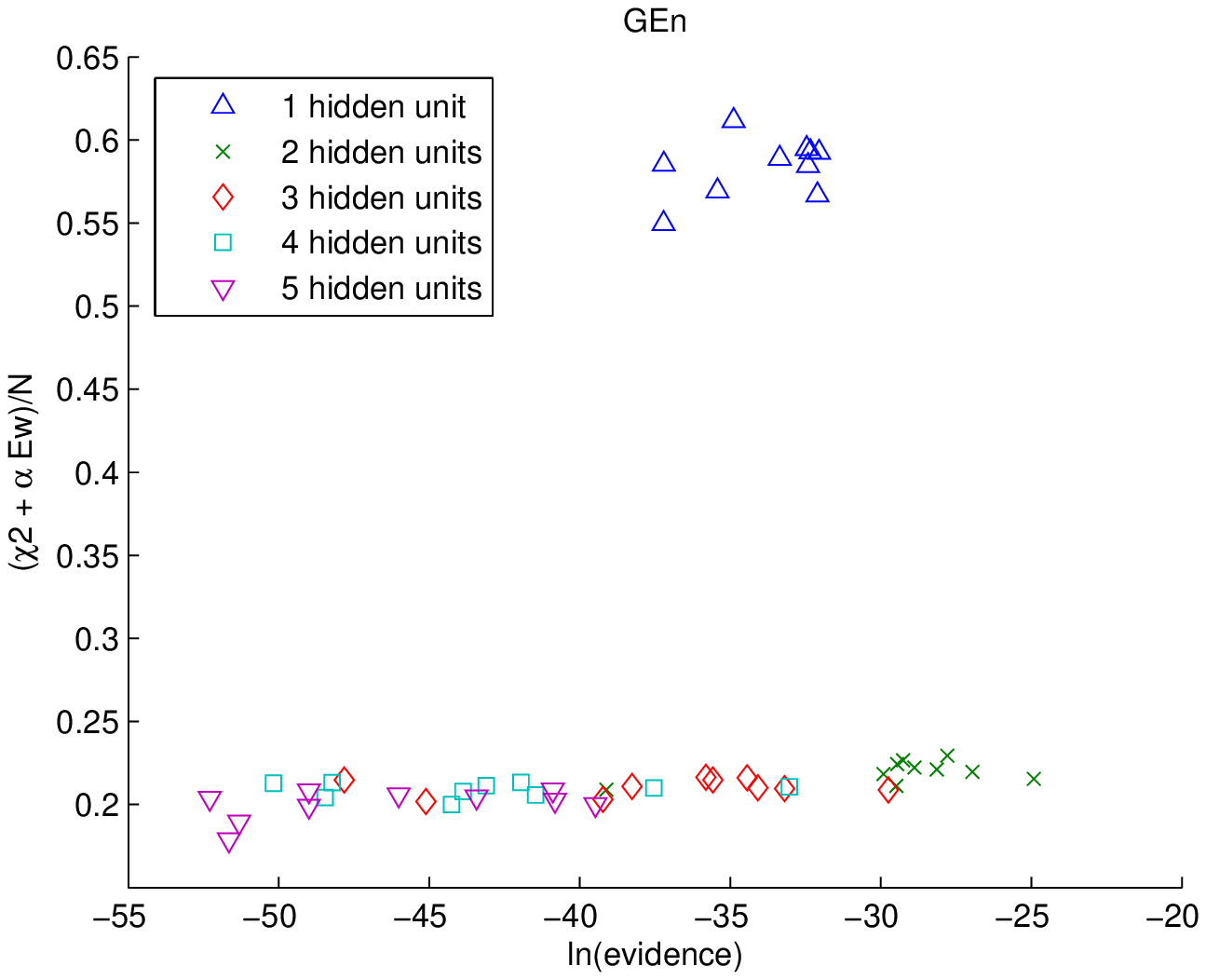, width=0.5\textwidth}{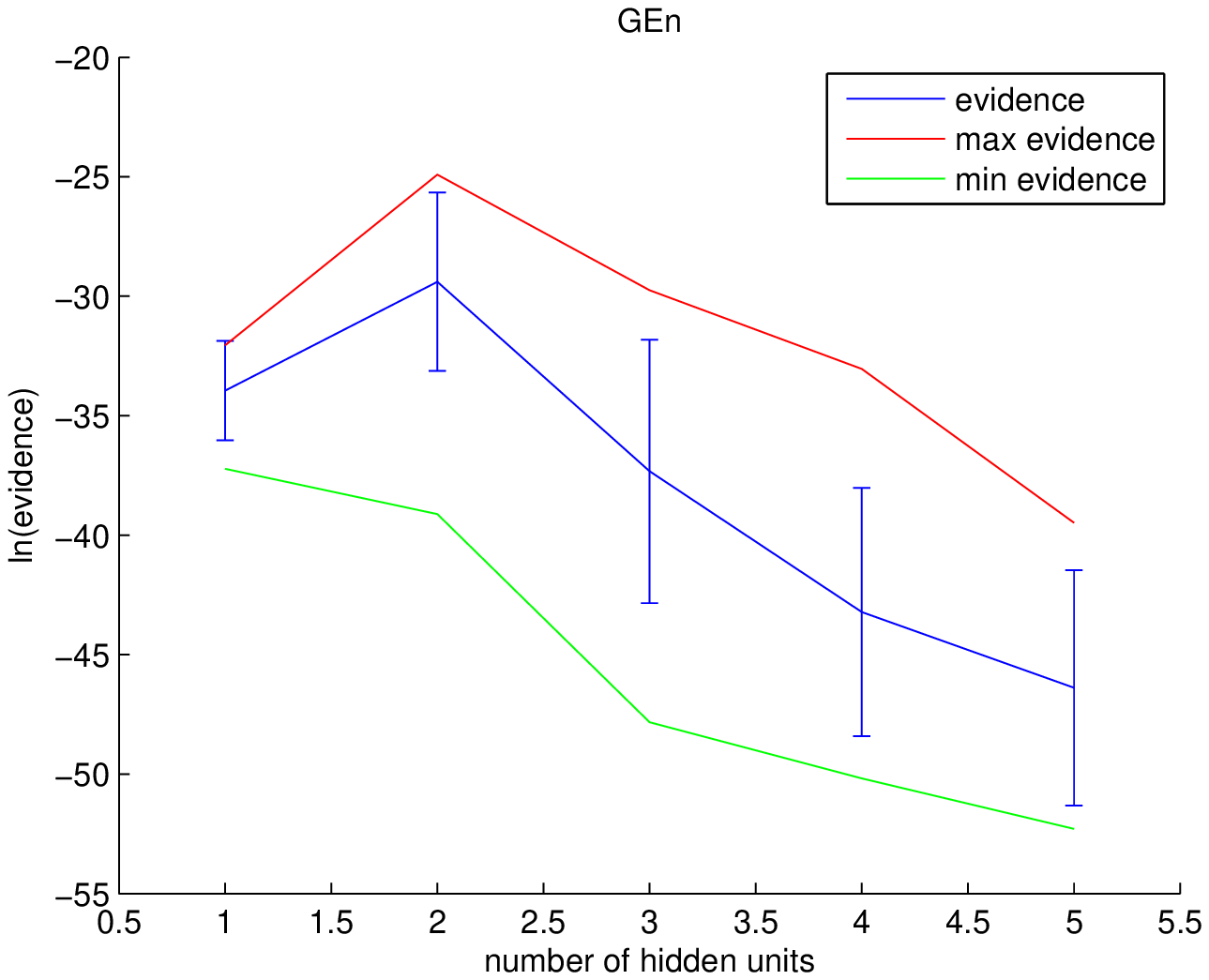, width=0.5\textwidth}{
The total error, $S(\vec{w}_{MP})$, as a function of $\ln \mathcal{P}\left(\mathcal{D}\right|\left. M \right)$ (ln evidence). The evidence is computed for networks trained with the $G_{En}$ data. The results obtained for  networks with $M=1-5$ hidden units are shown. Single point represents the fit obtained for given starting weight configuration and particular network type. \label{Fig_evidence_GEn}
}
{
The dependence of
$\ln \mathcal{P}\left(\mathcal{D}\right|\left. M \right)$ on the number of hidden units. The evidence is computed for networks trained with the $G_{En}$ data. The maximal  and minimal values of $\ln \mathcal{P}\left(\mathcal{D}\right|\left. M \right)$ (for given network type) are plotted with the red and green lines respectively.  The mean of $\ln \mathcal{P}\left(\mathcal{D}\right|\left. M \right)$ over all acceptable solutions  is represented by the blue line.
\label{Fig_evidence_GEn2}
}
\EPSFIGURE[pos]{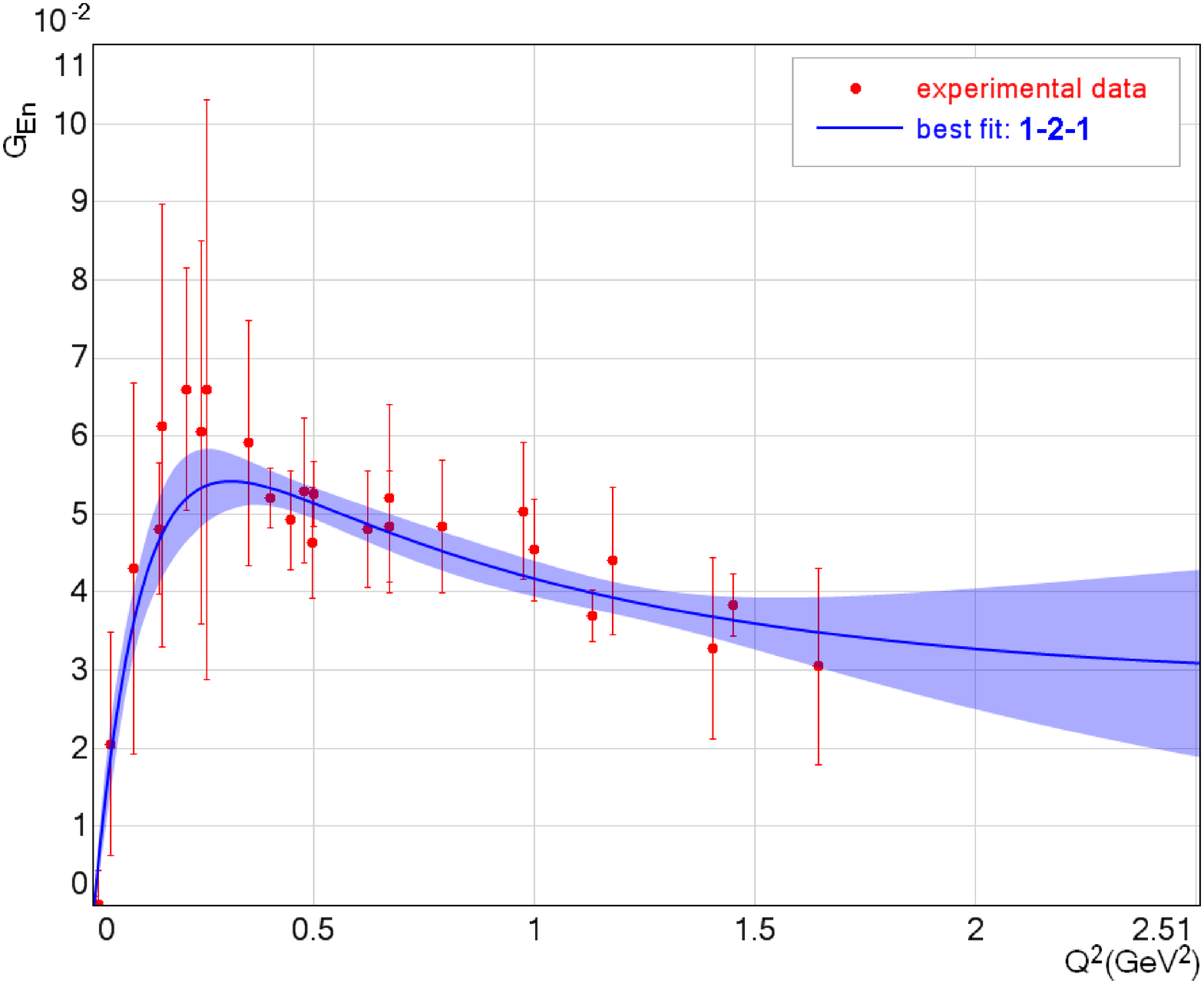, width=0.6\textwidth}{The best fit of $G_{En}$ data given by the 1-2-1 network. The blue area denotes fit uncertainty computed with Eq. \ref{1sigma_error}. The experimental data is the same as the one discussed in Ref. \cite{Alberico:2008sz}.
\label{Fig_GEn} }

The evidence is constructed in so called hierarchical approach. It is  a three level procedure. Applying Bayes' theorem the probability distribution for the weights parameters is constructed, then the probability distribution of the decay parameter $\alpha$, and eventually the evidence are evaluated.
\begin{eqnarray}
\mathcal{P}\left(\vec{w}\right|\left.  \mathcal{D},\alpha ,\mathcal{A}_M \right) &=& \frac{\mathcal{P}\left(\mathcal{D}\right|\left.\vec{w}, \alpha, \mathcal{A}_M \right)
\mathcal{P}\left(\vec{w}\right|\left. \alpha, \mathcal{A}_M \right)}{\mathcal{P}\left(\mathcal{D}\right|\left. \alpha, \mathcal{A}_M \right)} \to \\
\mathcal{P}\left(\alpha\right|\left. \mathcal{D}, \mathcal{A}_M \right)& =& \frac{\mathcal{P}\left(\mathcal{D}\right|\left. \alpha, \mathcal{A}_M \right) \mathcal{P}\left(\alpha\right|\left.  \mathcal{A}_M \right)}{\mathcal{P}\left(\mathcal{D}\right|\left.\mathcal{A}_M \right)}\to \\
\mathcal{P}\left(\mathcal{A}_M\right|\left.  \mathcal{D}\right) &=&\frac{\mathcal{P}\left(\mathcal{D}\right|\left. \mathcal{A}_M \right) \mathcal{P}(\mathcal{A}_M)}{\mathcal{P}(\mathcal{D})}.
 \end{eqnarray}
Below the short description of the Bayesian approach is presented.

\subsection*{1. Constructing the weight parameter distribution}

The probability distribution for the neural network weights is built, assuming that regularization parameter $\alpha$  is fixed:
\begin{equation}
\label{posterior_probability_w}
\mathcal{P}\left(\vec{w}\right|\left.  \mathcal{D}, \alpha ,\mathcal{A}_M \right) = \frac{\mathcal{P}\left(\mathcal{D}\right|\left.\vec{w}, \alpha, \mathcal{A}_M \right)
\mathcal{P}\left(\vec{w}\right|\left. \alpha, \mathcal{A}_M \right)}{\mathcal{P}\left(\mathcal{D}\right|\left. \alpha, \mathcal{A}_M \right)},
\end{equation}
where $\mathcal{P}\left(\vec{w}\right|\left. \alpha, \mathcal{A}_M \right)$ is a prior probability distribution of weights, while $\mathcal{P}\left(\mathcal{D}\right|\left.\vec{w}, \alpha, \mathcal{A}_M \right)$
is the likelihood function. In the case of present analysis the likelihood function is given by the $\chi^2$ function, namely:
\begin{equation}
\mathcal{P}\left(\mathcal{D}\right|\left.\vec{w}, \alpha, \mathcal{A}_M \right) = \frac{1}{Z_\chi} \exp[- \chi^2(\vec{w},\mathcal{D})], \quad
Z_\chi = \int d^N t \exp[- \chi^2(\vec{w},\mathcal{D})] =    \pi^{\frac{N}{2}}\prod_{i=1}^{N} \Delta t_i.
\end{equation}
The prior probability should be as general as possible. Indeed, there are plenty of possibilities (e.g. Laplacian or
entropy-based priors see discussion in Ref. \cite{Neal94}). We assume that every weight parameter is equally distributed according to a Gaussian distribution (with the zero mean and the variance of $1/\sqrt{\alpha}$)
\begin{equation}
\label{prior_alpha}
\mathcal{P}\left(\vec{w}\right|\left. \alpha, \mathcal{A}_M \right) = \frac{1}{Z_w(\alpha)} \exp[- \alpha E_w], \quad
Z_w(\alpha) = \int d^W w  \exp[- \alpha E_w]= \left(\frac{2\pi}{\alpha}\right)^\frac{W}{2}
\end{equation}
(the arguments supporting above choice of the prior are presented in  Sec. \ref{section_prior}).
It gives the  probabilistic interpretation for the regularization function $E_w$ defined in the previous section
(see Eq. \ref{regularyzator_definition}). Then we see  that:
\begin{eqnarray}
\label{PDalpha}
\mathcal{P}\left(\mathcal{D}\right|\left. \alpha, \mathcal{A}_M \right) &=& \int d^W w
\mathcal{P}\left(\mathcal{D}\right|\left. \vec{w}, \alpha, \mathcal{A}_M \right) \mathcal{P}\left(\vec{w}\right|\left. \alpha, \mathcal{A}_M \right) = \frac{Z_M(\alpha)}{Z_\chi  Z_w(\alpha)}, \\
Z_M(\alpha) &=& \frac{(2\pi)^{\frac{W}{2}}}{\sqrt{|A|}}\exp\left[-\chi(\vec{w}_{MP})-\alpha E_w(\vec{w}_{MP})\right]. \end{eqnarray}
The last integral was computed by expanding the error function up to the Hessian term:
\begin{equation}
\label{hessian_approximation}
S(\vec{w},\mathcal{D}) =  S(\vec{w}_{MP},\mathcal{D}) + \frac{1}{2}
(\vec{w} - \vec{w}_{MP})^T A (\vec{w} - \vec{w}_{MP}),
\end{equation}
where $\vec{w}_{MP}$ is the vector of weights which minimizes $S(\vec{w},\mathcal{D})$ (maximizes the posterior probability (\ref{posterior_probability_w})).

The Hessian matrix reads
\begin{eqnarray}
A_{ij} &=& \nabla_i \nabla_j \left. S \right|_{\vec{w} = \vec{w}_{MP}} = \nabla_i \nabla_j \chi^2(\vec{w},\mathcal{D}) + \alpha \delta_{ij} \\
&=&  2\sum_{k=1}^N \left[  \frac{\nabla_i y(x_k,\vec{w}_{MP}) \nabla_j y(x_k,\vec{w}_{MP})}{\Delta t_k^2} + \frac{\left(y(x_k,\vec{w}_{MP}) - t_k \right)}{\Delta t_k^2} \nabla_i \nabla_j y(x_k,\vec{w}_{MP}) \right] + \alpha \delta_{ij}.\nonumber \\
\label{hessian_def}
\end{eqnarray}
We compute the full Hessian matrix \cite{bishop_hessian}. Usually the double differential term in (\ref{hessian_def}) is neglected,
which is  a good approximation only at the minimum. Taking into account full Hessian plays a crucial role in optimizing $\alpha$ parameter, as it will become clear below.

The network response uncertainty $\Delta y $ is defined by the variance:
\begin{equation}
(\Delta y(x))^2 = \int d^W w \left[y(x,\vec{w}) - \left<y(x)\right>  \right]^2
\mathcal{P}\left(\vec{w}\right|\left. \alpha, \mathcal{D}, \mathcal{A}_M \right).
\end{equation}
In the first approximation it is expressed by the covariance matrix, i.e. inverse of the Hessian matrix:
\begin{equation}
\label{1sigma_error}
(\Delta y(x))^2 = (\nabla y(x,\vec{w}_{MP}))^T A^{-1} \nabla y(x,\vec{w}_{MP}).
\end{equation}
In Appendix A the covariance matrices obtained for every considered problem are presented.

\subsection*{2. Constructing $\alpha$ the distribution of the parameter $\alpha$ }

The $\alpha$ parameter is established by applying the so-called \textit{evidence approximation} \cite{MacKay92a,MacKay92b,Gull88b},
the method, which is equivalent to \textit{type II maximum likelihood} in conventional statistics.

The Bayes' rule  leads to:
\begin{equation}
\label{posterior_alpha}
\mathcal{P}\left(\alpha\right|\left. \mathcal{D}, \mathcal{A}_M \right) = \frac{\mathcal{P}\left(\mathcal{D}\right|\left. \alpha, \mathcal{A}_M \right) \mathcal{P}\left(\alpha\right|\left.  \mathcal{A}_M \right)}{\mathcal{P}\left(\mathcal{D}\right|\left.\mathcal{A}_M \right)},
\end{equation}
where the $\mathcal{P}\left(D\right|\left. \alpha  , \mathcal{A}_M \right)$ has been obtained in the previous section (see Eq. \ref{PDalpha}).

We are searching for the $\alpha_{MP}$ parameter,  i.e.  the one which maximizes the prior probability
(\ref{posterior_alpha}). It can be shown that in the \textit{Hessian approximation} it is given by the solution of the equation:
\begin{equation}
\label{alpha_MP_parameter}
 2 \alpha_{MP}E_w(\vec{w}_{MP}) =    \sum_{i=1}^W \frac{\lambda_i}{\lambda_i + \alpha_{MP}} \equiv \gamma,
\end{equation}
where $\lambda_i$'s are eigenvalues of the  matrix $ \nabla_n \nabla_m \left. \chi^2 \right|_{\vec{w} = \vec{w}_{MP}}$.  In practice, the
eigenvalues depend on $\alpha$, therefore to get a proper $\alpha_{MP}$ the $\alpha$ parameter is iteratively changed during
the training process i.e.:
\begin{equation}
\label{alpha_update_rule}
\alpha_{k+1} = \gamma(\alpha_{k})/2E_w(\vec{w}).
\end{equation}
The iteration procedure fixes in the optimal way the $\alpha$ parameter. The typical dependence of $\alpha_k$ on the iteration step is presented in Fig. \ref{Fig_alpha_iter}. In Sec. \ref{section_prior} it is shown that the choice of the initial $\alpha$ value has a small impact on the final results.

At the end of the training procedure one can approximate (\ref{PDalpha}) as follows:
\begin{equation}
\mathcal{P}\left(\mathcal{D}\right|\left. \ln\alpha  , \mathcal{A}_M \right) =
\mathcal{P}\left(\mathcal{D}\right|\left. \ln\alpha_{MP}  , \mathcal{A}_M \right)\exp
\left[-\frac{(\ln\alpha - \ln \alpha_{MP})^2}{2\sigma^2_{\ln\alpha}} \right],
\end{equation}
where in the \textit{Hessian approximation} $\sigma_{\ln\alpha} \approx {2}/{\gamma}$.

\subsection*{3. Constructing the evidence}
The evidence for given model is defined by denominator of (\ref{posterior_alpha}).
If one assumes the uniform prior distribution of
$\ln \alpha$ parameter\footnote{ It is the consequence of the fact that $\alpha$ is the scale parameter.} on some large $\ln \Omega$ region then the evidence can be approximated by:
\begin{equation}
\mathcal{P}\left(\mathcal{D}\right|\left.\mathcal{A}_M \right) \approx
\mathcal{P}\left(\mathcal{D}\right|\left.\alpha_{MP}, \mathcal{A}_i \right)\frac{2\pi\sigma_{\alpha}}{\ln\Omega}.
\end{equation}
The $\ln\Omega$ is a constant which is the same for the all  hypotheses.

The ln of evidence (we show only model independent terms) reads
\begin{equation}
\label{log_of_evidence}
\ln \mathcal{P}\left(\mathcal{D}\right|\left. \mathcal{A}_M \right)  \approx  -\chi^2(\vec{w}_{MP}) - \alpha_{MP} E_{w}(\vec{w}_{MP})
-\frac{1}{2}\ln |A| + \frac{W}{2}\ln \alpha_{MP} -\frac{1}{2}\ln \frac{\gamma}{2}.
\end{equation}
The first term  in the above expression, $-\chi^2(\vec{w}_{MP})$, (usually of low-value for simple models) is the misfit of the approximated data, while the next four terms
constitute the so called Occam factor, which penalizes the complex models. Since in this work we consider only the networks of
type 1-M-1 (only one hidden layer) in the rest of the paper we  will denote the evidence $\mathcal{P}\left(\mathcal{D}\right|\left. \mathcal{A}_M \right)$ by $\mathcal{P}\left(\mathcal{D}\right|\left. M \right)$.

\subsection{Prior Function}
\label{section_prior}

   We have already mentioned that the various possible prior distributions are considered in the literature \cite{Williams}. In this analysis the likelihood function  is given by $\chi^2$ distribution, which  has a Gaussian probabilistic interpretation. Therefore it seems to be reasonable to assume that the weight parameters distribution should also be  described  by the Gaussian-like prior function. Additionally we assume, without losing the generality, that:
   \begin{itemize}
   \item   negative, and positive values of the weight parameters are equally likely;
   \item at the beginning of the learning procedure the weight parameters are independent;
   \item  small\footnote{For the networks with the sigmoid  activation functions the non-trivial smooth functional parametrization are described by the low $|w_i|$ weights.} weight values are more likely than the large values.
   \end{itemize}
   Then the Gaussian-like prior distribution  can have a form:
 \begin{equation}
 \mathcal{P}\left(\vec{w}\right|\left. \alpha, \mathcal{A}_M \right) \sim  \exp\left[- \frac{1}{2}\sum_{i=1}^W \alpha_i w_i^2\right].
 \end{equation}
 Notice that every $w_i$ parameter has its own $\alpha_i$ regularization parameter. As it was mentioned in the previous section the $\alpha$ is the so-called scale parameter. The number of the scale parameters can be reduced if the symmetry property of the given network architecture is taken into account.  The network of the type 1-M-1 has: $M$ hidden weights; $M$ corresponding  bias weights and $M+1$ linear weighs (output weights + one bias parameter). The permutation between the hidden units does not change the network functional type. Permuting two hidden units  is realized by exchange between  the weight parameters of the same type (hidden, bias and linear weights). This symmetry property allows us to reduce the number of $\alpha$'s to three independent scale parameters:
 \begin{itemize}
 \item $\alpha_{h}$ for the hidden weights;
 \item $\alpha_{b}$ for bias weights (in hidden layer);
 \item $\alpha_{l}$ for  linear weights in the output layer.
 \end{itemize}
Then the prior function reads
 \begin{equation}
 \label{prior_3_alpha}
 \mathcal{P}\left(\vec{w}\right|\left. \alpha, \mathcal{A}_M \right) \sim  \exp\left[- \frac{1}{2}\left( \alpha_{h} \sum_{i \in \, hidden} w_i^2 + \alpha_{b} \sum_{i \in \, bias}  w_i^2+ \alpha_{l} \sum_{i \in \, linear}  w_i^2 \right) \right].
 \end{equation}

We made an  effort to compare results which are obtained with both   (\ref{prior_alpha}) and (\ref{prior_3_alpha}) priors. It was observed that final results are very similar. Analogically as in the case of (\ref{prior_alpha}) prior the $\alpha_h$, $\alpha_b$ and $\alpha_l$ parameters were iteratively changed during the training procedure.  The typical results, obtained for the $G_{Mn}/\mu_n G_D$ and $G_{En}$ data sets,
are shown in Fig. \ref{fig_alpha_regula}. The differences between the final best fits are negligible. In the left column of the same figure we plot the dependence of the $S(\vec{w},\mathcal{D})$ on the iteration step. We see that the minimal value of the total error is almost the same for both prior functions. For both cases the training  started from the same initial weight configuration.

All above seem to justify  the simplest choice of the prior function, namely the one given by Eq. \ref{prior_alpha}. Nevertheless, it may happen  that for more complex data then we discuss, the results will significantly depend on
 prior assumptions. In such case the Bayesian framework  can be used to indicate the best prior function.

Eventually, we discuss the  dependence of the final results on the initial $\alpha_0$ value.
 We considered several initial values of $\alpha_0$ (see Table \ref{table_alpha_distance_GMp}).
 After training we noticed that  the choice  of the initial  $\alpha_0$ had  a small
 impact on  the final $\alpha_{MP}$ value (see Fig \ref{Fig_alpha_iter}) as well as  the fits.
 It is shown in Table \ref{table_alpha_distance_GMp} where the relative distances, in the weight space, between  fits are presented. Notice that the only one solution computed for  $\alpha_0=1$ is out of others.
\TABULAR{|c| l |l|l|l|l|}{
\hline
$\alpha_0$	& 0.0001	& 0.001	& 0.01	& 0.1	& 1 \tt \\
\hline
0.0001 &	    0	    &   0.0925   & 	0.0196 	      & 0.8048  &	 14.8847 \tt \\
0.0010 &	    0.0925 	&   0	     &  0.0748  	  &  0.8921 & 	 14.9641 \tt \\
0.0100 &	    0.0196 	&   0.0748   &	0	          &  0.8214 &	  14.8965 \tt \\
0.1000 &	    0.8048 	&   0.8921 	 &  0.8214        &	0	    &      14.2768 \tt \\
1.0000 &	   14.8847  &	 14.9641 &	   14.8965    &	14.2768	&   0 \tt \\
	\hline
}{\label{table_alpha_distance_GMp} The distance $d(\vec{w}_{1}, \vec{w}_2) = \sqrt{\sum_{i=1}^W (w_{1i} - w_{2i})^2}$ between fits obtained for various initial $\alpha_0$ values. The computations are done for the $G_{En}$ data for the network of 1-2-1 type.}

It is  worth to mention that decreasing the $\alpha_0$ parameter  can be understood as enlarging the effective prior domain. For the final analysis we set $\alpha_0 = 0.001$.

In this section we have demonstrated  that  our results weakly depend on the prior assumptions.
It has been also shown   that it is relatively easy to construct the prior function if the symmetry properties of network are taken into consideration. Usually, it is not the case in the conventional form-factor data analysis, where the ad-hoc parametrizations are discussed. The typical phenomenological parametrization has no straightforward symmetries. As an example consider the function \cite{Kelly:2004hm,Arrington:2007ux}:
\begin{equation}
G(Q^2) = \frac{a_0 + a_1 Q^2 + a_2 Q^4}{b_0 + b_1 Q^2 + b_2 Q^4 + b_3 Q^6 + b_4 Q^8}.
\end{equation}
Constructing  the prior function for above form-factor parametrization seems to be more complicated than in the ANN case. One can  postulate the values of the ratios
$a_0/b_0$ and  $a_2/b_4$, which describe the low and high $Q^2$ behavior of the FF. However,
the rest of parameters, which  seem to model the intermediate $Q^2$ region,  can have any arbitrary values. Therefore building the prior distribution for above FF would require an extra phenomenological and theoretical knowledge.

\TABULAR{|c| c |c|}{
	\hline	\tt  error of 	&  &   \tt \\
		\tt artificial  	& $G_{Mn}/\mu_n G_D$ & error \tt \\
		\tt point  ($\Delta t$)	&  & \tt \\
\hline
\multicolumn{3}{|c|}{$Q^2$=0}\\
\hline
\tt 1.0000  &       1.01809  &  0.03258 \tt  \\
\tt 0.1000  &       1.01633  &  0.03097 \tt \\
\tt 0.0100  &       1.00198  &  0.00947 \tt \\
\tt 0.0010  &       1.00002  &  0.00075 \tt \\
\tt 0.0001  &       1.00000  &  0.00009 \tt \\
	\hline
\multicolumn{3}{|c|}{$Q^2$=0.1}\\
\hline
\tt 1.0000   &      0.96920 &  0.00700 \tt \\
\tt 0.1000   &      0.96893  & 0.00696 \tt \\
\tt 0.0100   &      0.96709 &  0.00642 \tt \\
\tt 0.0010   &      0.96650 &  0.00624 \tt \\
\tt 0.0001   &      0.96986 &  0.00599 \tt \\
\hline
\multicolumn{3}{|c|}{$Q^2$=1.0}\\
\hline
\tt 1.0000      &   1.03657  & 0.00792\tt \\
\tt 0.1000      &   1.03669  & 0.00797\tt \\
\tt 0.0100      &   1.03720  & 0.00815\tt \\
\tt 0.0010      &   1.03778  & 0.00813\tt \\
\tt 0.0001      &   1.03575  & 0.00773\tt \\
\hline
}{\label{table_artificial_GMn} Dependence of $G_{Mn}/\mu_n G_D$ and its uncertainty (computed for $Q^2=$0, 0.1, and 1)  on the $\Delta t$ of the artificial point   added at $Q^2=0$.   }

\section{Form-Factor Fits}
\label{section_results}

\subsection{Data}
We consider the electric and magnetic proton and neutron
form-factor data. The electric and magnetic nucleon form-factors are defined as follows:
\begin{eqnarray}
\label{Sachs_Form_Factors_def}
G_{Mp,n}(Q^2)   &=& F_1^{p,n}(Q^2)  +F_2^{p,n}(Q^2), \\
G_{Ep,n}(Q^2)   &=& F_1^{p,n}(Q^2) - \frac{Q^2}{4M^2} F_2^{p,n}(Q^2),
\end{eqnarray}
where:
\begin{equation}
G_{Mp,n}   = \mu_{p,n}, \quad G_{Ep} = 1, \quad G_{En}=0.
\end{equation}
The experimental data is usually normalized to the dipole form-factor $G_D=1/(1 +Q^2/0.71)^2$.

The electric $G_{Ep}$ and magnetic $G_{Mp}$ proton FF data have been  obtained via Rosenbluth
separation technique from elastic $ep$  scattering \cite{Qattan:2004ht}. Additionally since the beginning of nineties of last century the
measurement of the form-factor ratio $\mu_p G_{Ep}/G_{Mp}$ in the spin dependent elastic $ep$ scattering
have been performed \cite{Gayou:2001qt}.

It turned out that  systematic discrepancy exists between so-called \textit{Rosenbluth}  and \textit{polarization
transfer} $\mu_p G_{Ep}/G_{Mp}$ ratio  data. The difference
can be  explained when the two photon exchange effect (TPE) \cite{TPE_papers}
is taken into account (for review see \cite{Carlson:2007sp}).  Hence, a proper fit of the EM form-factors requires
to take into account the TPE correction \cite{Alberico:2008sz}. In this work  we consider the re-analyzed  (TPE corrected Rosenbluth)
$G_{Mp}/\mu_p G_D$ and $G_{Ep}/G_D$ data  (Tabs. 2 and 3 of Ref. \cite{Arrington:2007ux}).
However, to see the TPE effect we consider also the original, (called here \textit{old Rosenbluth data}) $G_{Mp}/\mu_p G_D$ \cite{Qattan:2004ht,proton_form_factor_data,the_rest_of_GMp_olddata} and $G_{Ep}/G_D$ \cite{Qattan:2004ht,proton_form_factor_data,the_rest_of_GEp_olddata} data sets\footnote{We used the JLab data-base \cite{JLab_data}.}.
The neutron form-factor data ($G_{En}$ and $G_{Mn}$) are obtained from the electron scattering off light nuclei
(deuteron \cite{BLAST:2008ha}, helium \cite{Xu:2002xc}). Since the complexity of nuclear target, getting nucleon form-factors is more demanding than in the case of the elastic $ep$  scattering. The ground and final states of the nucleon must be
properly described. In this analysis we consider the same $G_{En}$ and $G_{Mn}/\mu_n G_D$ data sets as in Ref. \cite{Alberico:2008sz}.

Let us mention that to obtain proper fits of the form-factors at $Q^2=0$ we added to every data set one artificial point, namely $(Q^2=0,t=1,\,\Delta t= 0.001 )$ for $G_{Mn}/\mu_n G_D$, $G_{Mp}/\mu_p G_D$ and $G_{Ep}/G_D$ data sets, and $(Q^2=0,t=0,\,\Delta t= 0.001 )$ for $G_{En}$ data set.  This constraints have an effect on the final fit value and the uncertainty only in the close surrounding of the added point, as it is shown in Table \ref{table_artificial_GMn}, where we present how the best fit values and its uncertainties depend on the artificial point error. We
present results for $G_{Mn}$ data but for other considered data sets we  got analogical conclusions. The $\Delta t$ value assigned to the additional point should be comparable to data uncertainties used in the network
training. We have found that using $\Delta t = 0.01$ and higher is not sufficient to
attract the fit to desired value at constraint point, while $\Delta t = 0.0001$ causes
numerical difficulties during the training since the point has dominant contribution to
the overall network error value.

\subsection{Numerical Procedure}
\label{section_numerical_procedure}
The numerical analysis was done with two independent neural network softwares (in order to cross-validate the results). One written by R.S. and P.P. \cite{sulej_network} and another, which has been developed by K.M.G. \cite{granet}.

The procedure for finding the best neural network model for each data set consists of the five major steps:
\begin{enumerate}
\item \label{first_step} the sequence of networks of different 1-M-1 type (1-1-1, 1-2-1, 1-3-1, ... etc.) is taken into consideration;
\item \label{second_step} for each network of 1-M-1 type  the sample  of networks with randomly  initiated weights is trained;
\item among the networks obtained in the previous step,  ten networks with the lowest total error are selected for further analysis, see Sec. \ref{section_training_networks} and the $S(\vec{w}_{MP},\mathcal{D})$ distributions  shown in Figs. \ref{Fig_GEP_chi2} and \ref{Fig_GMP_chi2};
\item the network (from the step above) with the highest evidence is chosen as the best fit candidate for  given network type;
\item the best fits obtained for every network type are compared; the one with the highest   evidence is chosen to represent the data.
    \end{enumerate}
Let us remind that in the second step the large number (from 150 to 1300) of networks in the sample (as it is explained in Sec. \ref{section_training_networks}) is considered in order to find the solutions which maximizes the posterior probability for the given model.

The procedure for the single network training is as follows (see Fig. \ref{Fig_schemat_blokowy}):
\begin{itemize}
\item initialize the network weights as  small random values;
\item initialize the regularization factor (Eq. \ref{regularyzator_definition}), in this analysis $\alpha_0 = 0.001$;
\item perform the network training iterations, according to the \textit{Levenberg-Marquardt},  \textit{quick-prop}, or \textit{rprop} algorithms;
\item calculate the updated regularization factor $\alpha_{k+1}$ (Eq. \ref{alpha_update_rule}) every 20 iterations of the training algorithm; eigenvalues of Hessian matrix below $10^{-6}$ are rejected from the evaluation of $\gamma(\alpha_k)$ (Eq. \ref{alpha_MP_parameter});
\item calculate the network output (Eq. \ref{nn_output}) and uncertainty (Eq. \ref{1sigma_error}) values for the given range of $Q^2$ values;
\item calculate the ln of   evidence (Eq. \ref{log_of_evidence})
\end{itemize}

Eventually, we will shortly highlight the major differences between the  NNPDF approach
and the one presented in this article.

In this work we consider the sequence of networks with graded number of hidden units.
With the help of the Bayesian framework the best solution is chosen.
The NNPDF group considers one particular network architecture (2-5-3-1 type) to fit the data \cite{Ball:2008by}.
But some discussion of the dependence of results on the network architecture is presented.

The NNPDF group prepares the sample of the networks. Each network from the sample is trained with the artificial data
which is Monte Carlo generated from the original measurements.
Then the best fit and its uncertainty are obtained as an average and standard deviations computed over the sample. In this work every network is always trained with the original data set. Nevertheless the large sample of networks of given type is prepared but in order to find the architecture  and the weights which maximize the evidence. The network response uncertainty is computed from the covariance matrix (Eq. \ref{1sigma_error}).

Both approaches deal with the over-fitting problem  but in different ways.
The NNPDF applies the early stopping in the training (cross-validation algorithm is imposed).
Whereas we consider the regularization penalty term in the error function, which is optimized by the Bayesian procedure.
Hence the approach we apply  does not require validation of the solutions by comparing with the  test data set.

\subsection{Numerical Results}
The numerical procedures described in the previous section were applied to  all (six) the data sets.
We consider networks with $M=1-5$ hidden units for $G_{Mn}$, $G_{En}$, and $G_{Ep}$ data and with
$M= 1- 6$ for the $G_{Mp}$ data.  The evidence quantitatively classifies the networks i.e.  the most suitable network architecture for representing the data is indicated by the maximum of the evidence.   Notice that the optimal way to deal with these results would be taking an average over all solutions weighted by the evidence. However, in all problems considered here we obtained clear signal (a peak at the evidence) for particular solution. It allowed us to neglect the contribution from networks of other size.
\DOUBLEFIGURE[t]{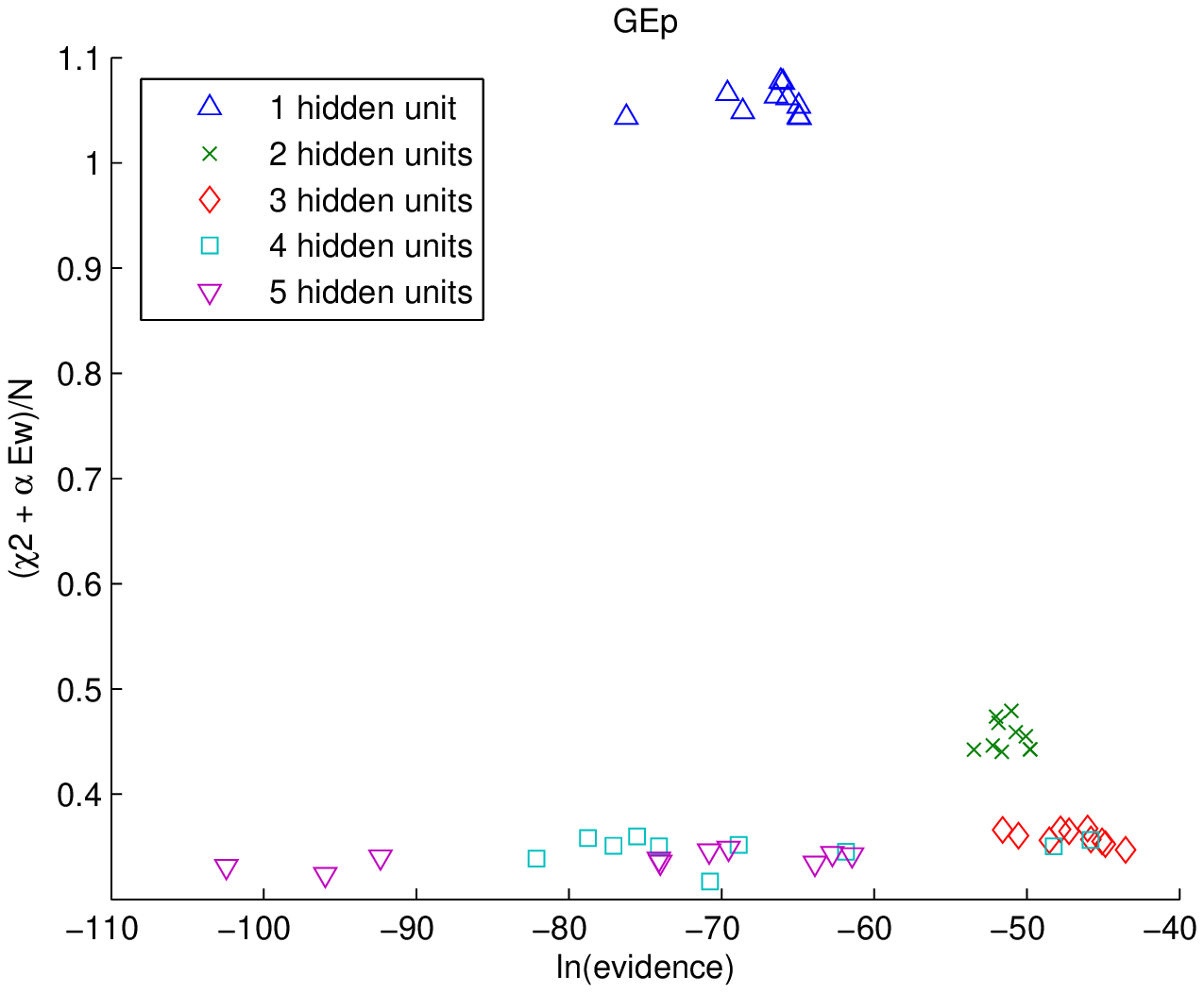,width=0.5\textwidth}{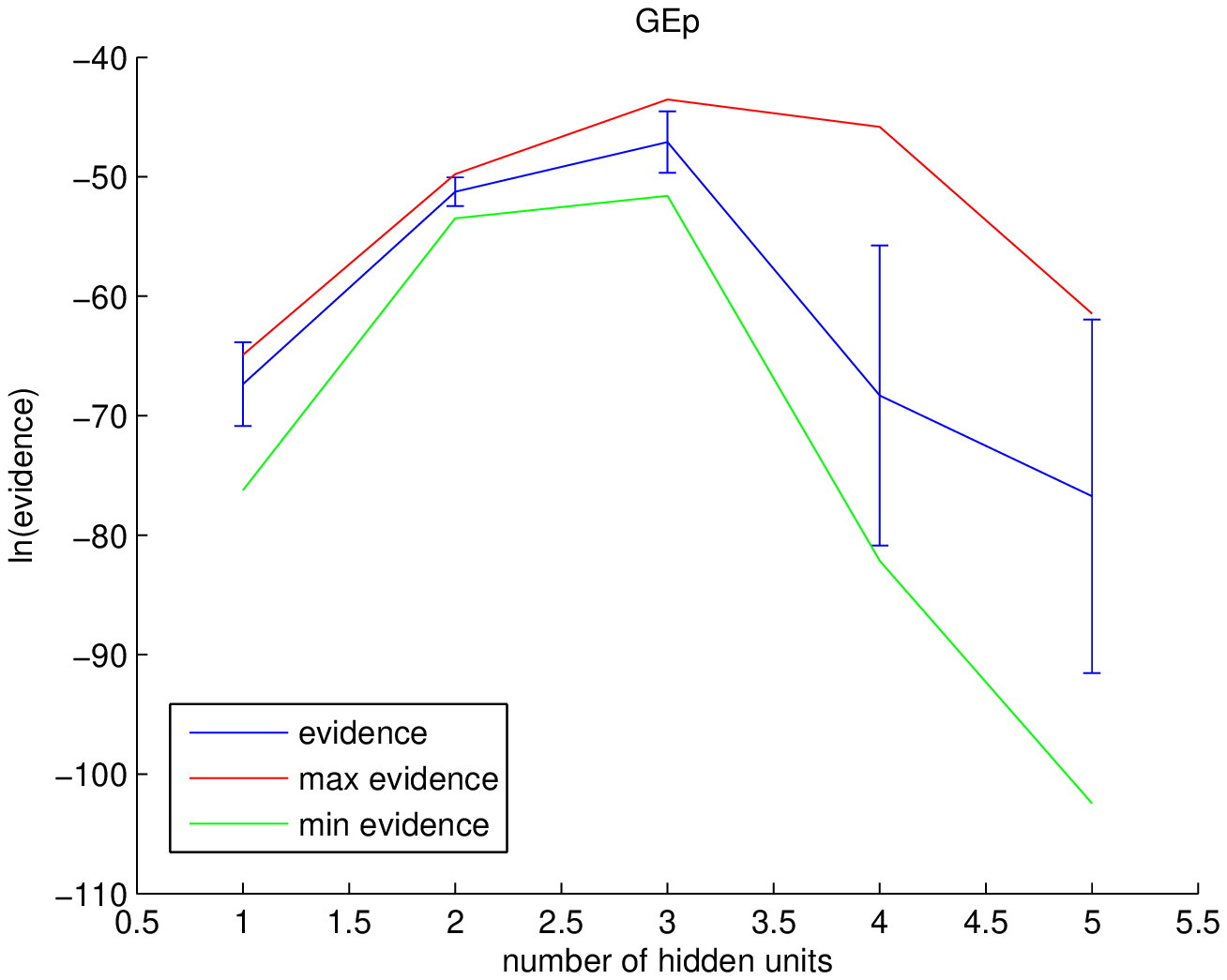,width=0.5\textwidth}{The total error, $S(\vec{w}_{MP})$, as a function of $\ln \mathcal{P}\left(\mathcal{D}\right|\left. M \right)$ (ln evidence). The evidence is computed for networks trained with the $G_{Ep}/G_D$ data. The results obtained for  networks with $M=1-5$ hidden units are shown. Single point represents the fit obtained for given starting weight configuration and particular network type. \label{Fig_evidence_GEp}}{The dependence of
$\ln \mathcal{P}\left(\mathcal{D}\right|\left. M \right)$ on the number of hidden units. The evidence is computed for networks trained with the $G_{Ep}/G_D$ data. The maximal  and minimal values of $\ln \mathcal{P}\left(\mathcal{D}\right|\left. M \right)$ (for given network type) are plotted with the red and green lines respectively.  The mean of $\ln \mathcal{P}\left(\mathcal{D}\right|\left. M \right)$ over all acceptable solutions  is represented by the blue line.
\label{Fig_evidence_GEp2}}
\DOUBLEFIGURE[t]{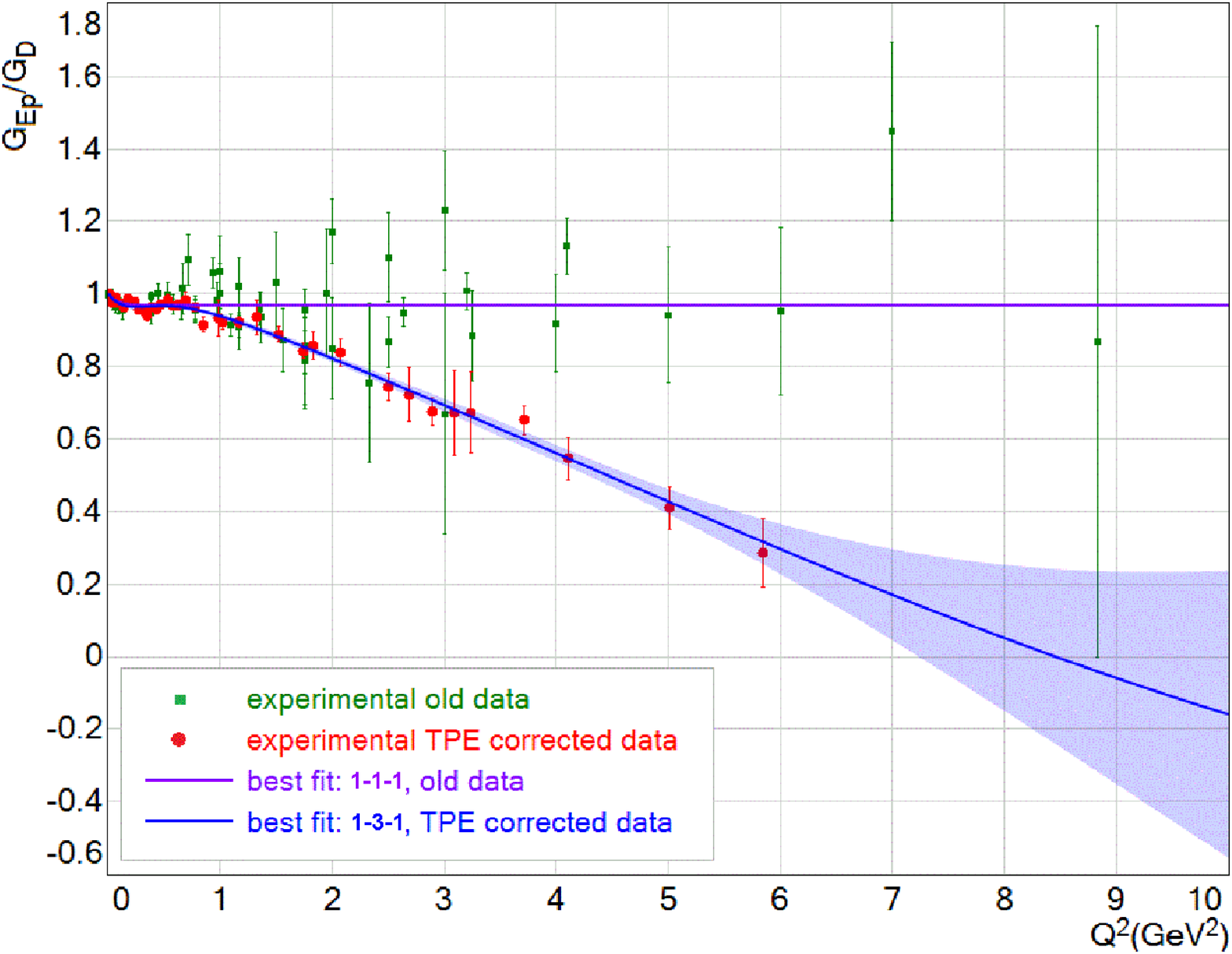,width=0.5\textwidth}{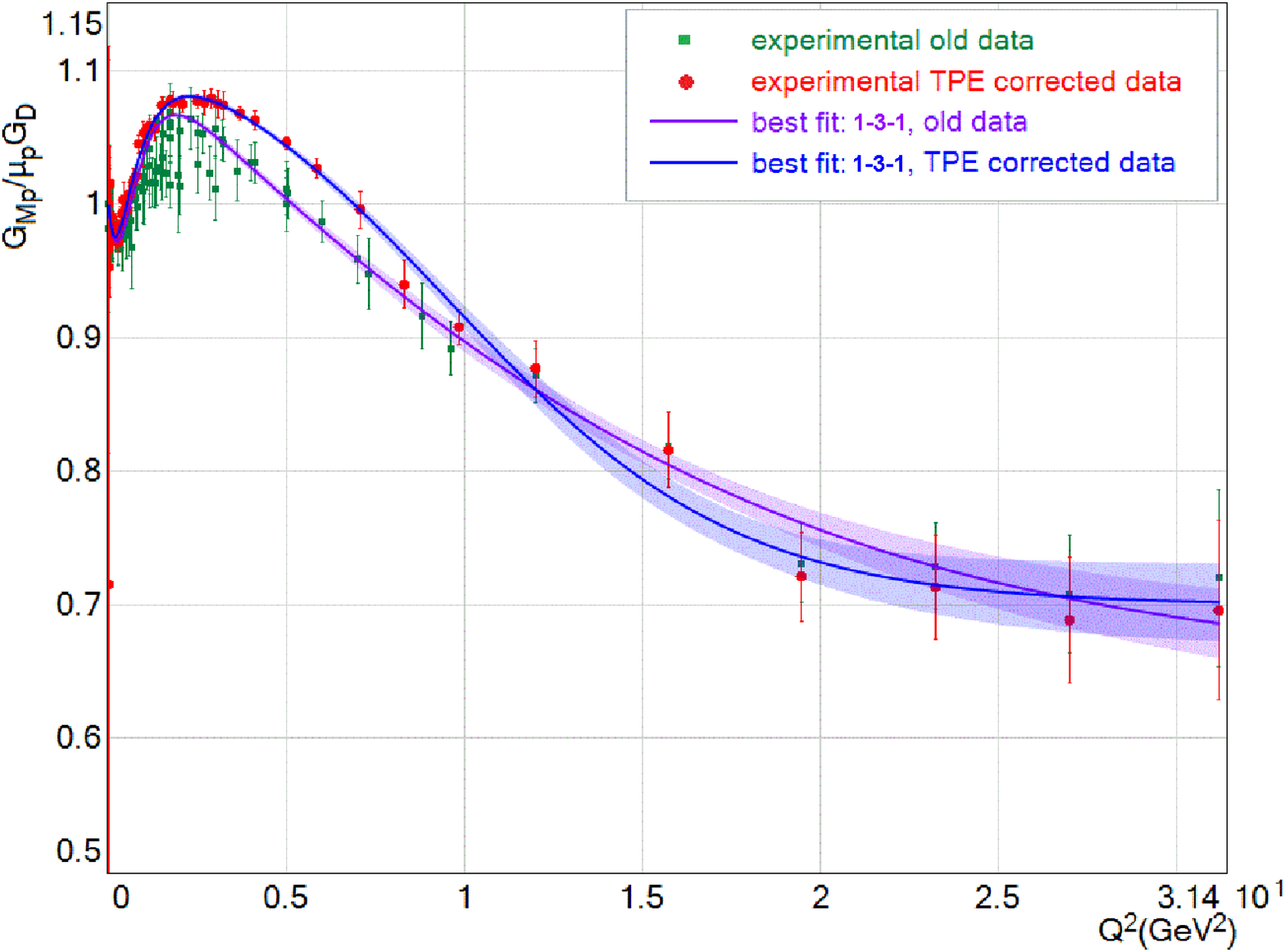,width=0.5\textwidth}{The best fit of $G_{Ep}/G_D$  data. The fit to TPE corrected data is given by 1-3-1 network (blue line), the data (red points) is taken from \cite{Arrington:2007ux}. The fit to ''old Rosenbluth data'' (green points) is given by 1-1-1 network (violet line),  the data  is taken from \cite{Qattan:2004ht,proton_form_factor_data,the_rest_of_GEp_olddata}. The fit uncertainty  is computed with Eq. 3.16.\label{Fig_GEp}}{The best fit of $G_{Mp}/\mu_p G_D$  data given by the 1-3-1 network.  The fit to TPE corrected data is given by 1-3-1 network (violet line), the data (red points) is taken  from \cite{Arrington:2007ux}. The fit to ''old Rosenbluth data'' (green points) is given by 1-1-1 network (violet line),  the data  is taken from  \cite{Qattan:2004ht,proton_form_factor_data,the_rest_of_GMp_olddata}. The fit uncertainty  is computed with Eq. 3.16.\label{Fig_GMp}}

We start the presentation of the numerical results by  the  discussion of the  $G_{Mn}/\mu_n G_D$ FF data. As it was described above, we consider a set of networks, which differ by number of hidden units $M$.  In Fig. \ref{Fig_evidence_GMn}  we show the scatter plot presenting the dependence of given network size on error function and log of evidence. One can notice that the networks 1-2-1 and 1-3-1 have the highest evidences, but the networks with $M=2$ hidden units  are not able to reproduce as low total error value as 1-3-1 networks. It is interesting also to mention that for $M\geq 3$ the total error slowly varies, i.e. increasing the number of the hidden units lowers  the total error by the minor amount. The clear indication for 1-3-1 network type is seen in Fig. \ref{Fig_evidence_GMn2}, where  only dependence of $\ln \mathcal{P}\left(\mathcal{D}\right|\left. M \right)$ on  $M$ is shown. In this figure we plot the maximal evidences obtained for given network type. However, in order to control the stability of numerical procedure we plot also the ln of evidence averaged over the networks around global minimum (solutions selected in step 3, Sec. \ref{section_numerical_procedure}), as well as the ln of the  minimal values of $\mathcal{P}(\mathcal{D}\left|\mathcal{M} \right.)$.
\DOUBLEFIGURE[t]{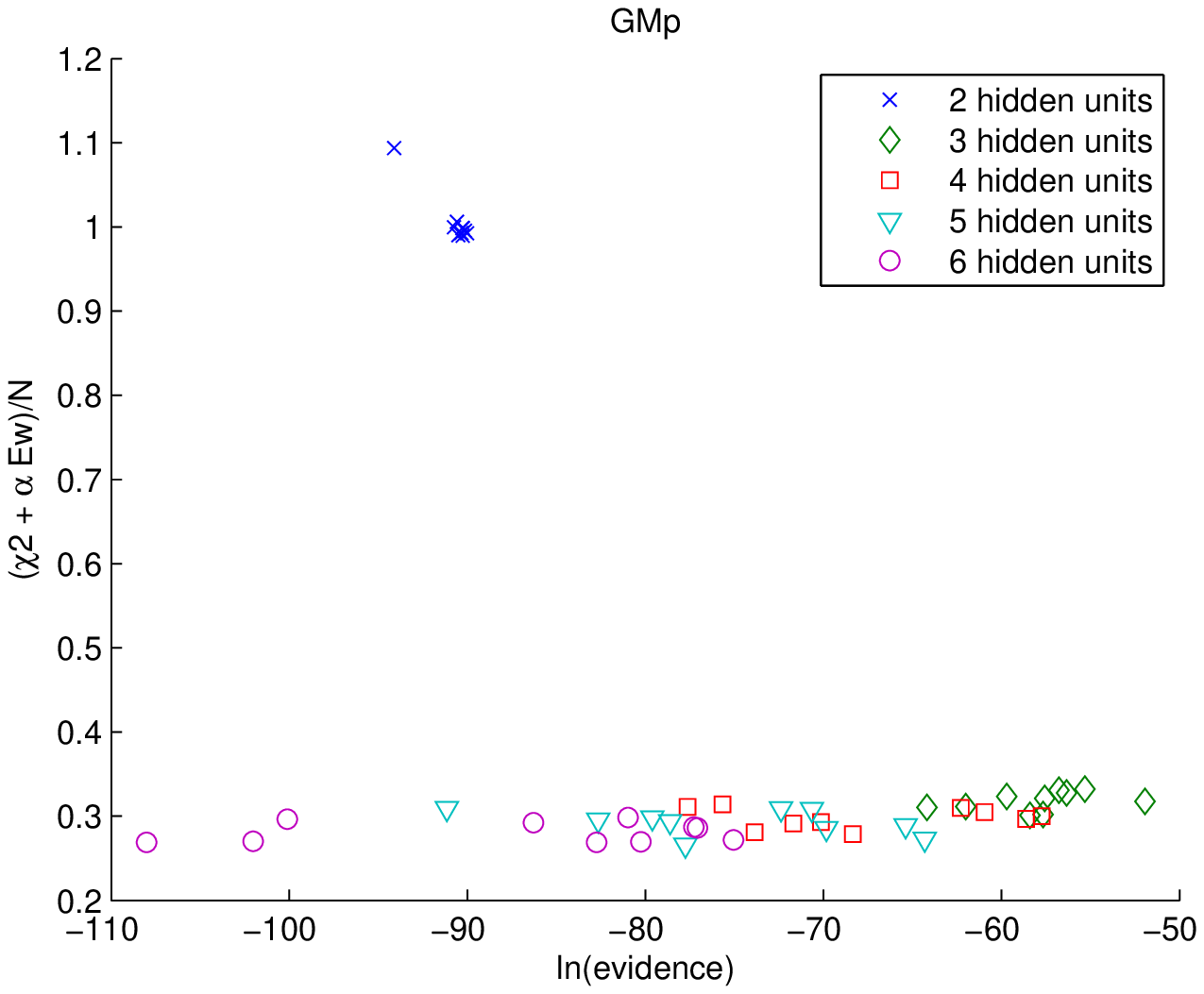,width=0.5\textwidth}{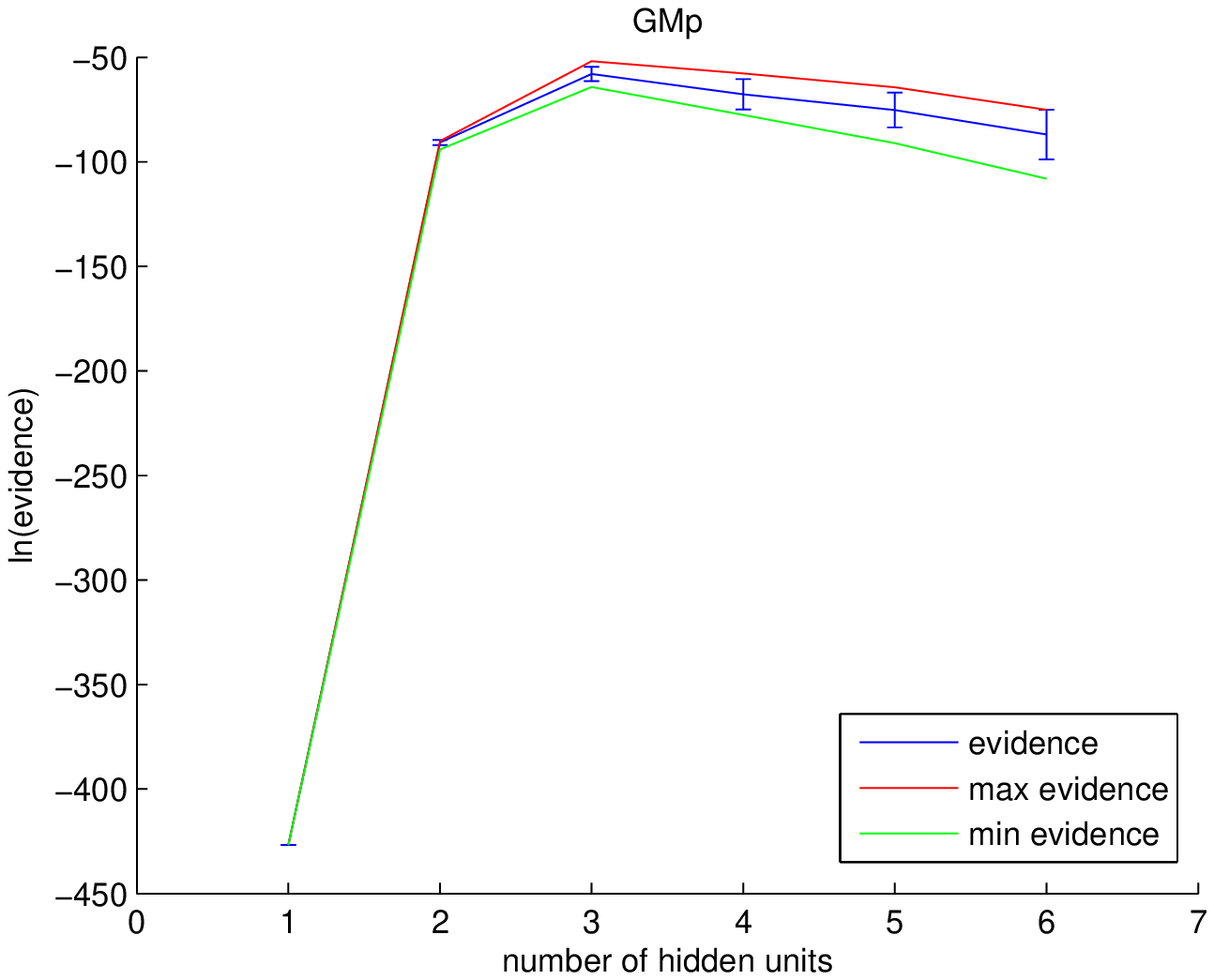,width=0.5\textwidth}{The total error, $S(\vec{w}_{MP})$, as a function of $\ln \mathcal{P}\left(\mathcal{D}\right|\left. M \right)$ (ln evidence). The evidence is computed for networks trained with $G_{Mp}/\mu_p G_D$ data. The results obtained for  networks with $M=1-6$ hidden units are shown. Single point represents the fit obtained for given starting weight configuration and particular network type. \label{Fig_evidence_GMp}}{The dependence of
$\ln \mathcal{P}\left(\mathcal{D}\right|\left. M \right)$ on the number of hidden units. The evidence is computed for networks trained with $G_{Mp}/\mu_p G_D$ data. The maximal  and minimal values of $\ln \mathcal{P}\left(\mathcal{D}\right|\left. M \right)$ (for given network type) are plotted with the red and green lines respectively.  The mean of $\ln \mathcal{P}\left(\mathcal{D}\right|\left. M \right)$ over all acceptable solutions  is represented by the blue line. \label{Fig_evidence_GMp2}}

All together suggest the network of type 1-3-1  (with the highest evidence) for the best fit of the $G_{Mn}$ data. The network output  is drawn in Fig. \ref{Fig_GMn}  together with the experimental data. The neural network response uncertainty is computed with (\ref{1sigma_error}) expression and shown in Fig. \ref{Fig_GMn2}.  In Fig.~\ref{Fig_GMn}  we plot also the best fits obtained for networks: 1-1-1, 1-2-1, 1-4-1 and 1-5-1. As could be expected increasing the number of hidden units makes the fit more flexible.

The electric neutron FF data ($G_{En}$) is analyzed in  the same way as the magnetic neutron one.
In Figs. \ref{Fig_evidence_GEn}, \ref{Fig_evidence_GEn2} and \ref{Fig_GEn} the plots of evidence and $G_{En}$ form-factor are shown. For $M=2$ we obtained the peak of the Occam's hill, what indicates 1-2-1 network architecture as the most representative parametrization.

The results for the electric and magnetic FF data are presented in Figs. \ref{Fig_evidence_GEp}, \ref{Fig_evidence_GEp2} and \ref{Fig_evidence_GMp}, \ref{Fig_evidence_GMp2}
(scatter and evidence plots) and Figs. \ref{Fig_GEp} and \ref{Fig_GMp} (form-factor plots). The network of type 1-3-1 is preferred
by  the both   electric and magnetic data sets. As it has been mentioned above we analyzed also the old form-factor data, which are not TPE corrected. It was obtained that the old $G_{Ep}$ prefers representation by the network of type 1-1-1. Hence, the old Rosenbluth $G_{Ep}/G_D$ data fit is almost linear constant function in $Q^2$. But the data seems to be not conclusive enough, so the Bayesian procedure leads to the simplest possible solution. On the other hand, it means that the old proton electric data does not show clear indication for deviation from the dipole form.

\subsection{Summary}
We have analyzed the form-factor data by the means of the artificial neural networks. The Bayesian approach has been adapted for the $\chi^2$ minimization and then applied to the data analysis. For every form-factor data set  sequence of neural networks  have been considered. The  Bayesian approach provided us with an objective criteria for choosing the most suitable form-factor parametrization (neural network) with the statistically optimal balance of the fit complexity and its uncertainty.   Therefore the resulting fits are unbiased and model independent. It has been demonstrated also that the final results weakly depend on the prior assumptions.

The approach  allowed to investigate objectively  the non-dipole deviations of the form-factors. It is interesting to mention that the   $G_{Ep}/G_D$, $G_{Mp}/\mu_p G_D$ as well as $G_{Mn}/\mu_n G_D$  form-factor data prefer  the same type (size) network 1-3-1.
The form-factor parametrizations, obtained in this analysis  can be easily applied to any phenomenological and experimental analysis.
Additionally, a part of the our software used in the analysis is available at \cite{sulej_network,granet}.

Presented method seems to be a promising statistical framework for studying and representing the experimental data. Especially, if the theoretical predictions are not able to reproduce measurements with desired accuracy, but the experimental data is sufficiently comprehensive to describe physical quantity by itself.

\section*{Acknowledgements}
K.M.G. thanks Carlo Giunti for inspiring discussions at the early stage of this project.

\appendix
\section{Analytical Formulae}

The two parametrizations of the form-factors have been obtained. The network of the type 1-2-1 representing $G_{En}$:
\begin{equation}
G_{En}(Q^2) = w_5 f_{act}(Q^2w_1 + w_2) + w_6  f_{act}(Q^2 w_3 + w_4) + w_7,
\end{equation}
and the network of the type 1-3-1, representing $G_{Mn}/\mu_n G_D$, $G_{Ep}/G_D$ and $G_{Mp}/\mu_p G_D$:
\begin{eqnarray}
G_{f}(Q^2)/g G_D &=& w_7 f_{act}(Q^2w_1 + w_2) + w_8  f_{act}(Q^2 w_3 + w_4) + w_9  f_{act}(Q^2 w_5 + w_6) + w_{10},\nonumber \\
 f=Mm, Ep, Mp, & &
\end{eqnarray}
where $g=1$ for proton electric form-factor and $g=\mu_{p,n}$ for the proton, neutron magnetic form-factors.  The activation function reads
\begin{equation}
f_{act}(x) = \frac{1}{1 + \exp(-x)}.
\end{equation}

The weights obtained for $G_{En}$:
\begin{equation}
\vec{w}^T_{MP}= \scriptsize{(10.19704, 2.36812, -1.144266, -4.274101, 0.8149924, 2.985524, -0.7864434})
\end{equation}
with the covariance matrix:
\begin{equation}
A^{-1}= \tiny{\pmatrix{
77182.936 & -76674.953 & 11320.149 & -976.911 & -59149.683 & -510.459 & 59023.698\cr
-141838.399 & 158041.683 & -17763.896 & 1808.806 & 121039.907 & 875.737 & -120845.155\cr
1007.74 & 1987.396 & 2153.904 & 94.542 & 1216.369 & 99.514 & -1244.23\cr
-881.971 & 1138.085 & 154.164 & 2325.543 & 841.299 & -6673.131 & -844.274\cr
-106233.935 & 117881.485 & -13555.199 & 1345.44 & 90325.25 & 660.27 & -90176.259\cr
-524.981 & -282.957 & -492.929 & -6713.68 & -132.119 & 19769.326 & 138.851\cr
106231.986 & -117915.687 & 13528.692 & -1347.504 & -90347.707 & -661.274 & 90199.073}
}.
\end{equation}
The weights obtained for $G_{Mn}/\mu_n G_D$:
\begin{equation}
\vec{w}^T_{MP}=\scriptsize{(3.19646, 2.565681, 6.441526, -2.004055, -0.2972361, 3.606737, -3.135199, 0.299523, 1.261638, 2.64747)}
\end{equation}
with the covariant matrix:
\begin{equation}
A^{-1}=\tiny{\pmatrix{
13019.47 & 5437.135 & 1625.832 & 2407.977 & 2421.111 & -9226.711 & -5508.625 & -466.761 & 11858.122 & -5018.992\cr
-110.632 & 2389.64 & -1419.064 & 1007.869 & 68.926 & 748.578 & -8134.262 & 132.414 & 1320.543 & 6692.985\cr
1186.145 & 1096.726 & 5283.129 & -2368.423 & -32.076 & 97.757 & 331.547 & -371.507 & -157.415 & 188.748\cr
2412.026 & 476.941 & -2382.018 & 2014.753 & 486.682 & -1841.165 & -1386.815 & 128.242 & 2393.05 & -961.438\cr
1688.083 & 877.17 & -114.146 & 433.604 & 445.447 & -1374.196 & -1269.078 & -43.863 & 2404.463 & -943.785\cr
-15867.205 & -7087.878 & -406.067 & -3161.81 & -3587.665 & 16599.626 & 7510.982 & 544.902 & -14185.447 & 4857.133\cr
9913.113 & -1376.838 & 7840.949 & -2871.111 & 1670.017 & -9080.441 & 18045.64 & -1025.567 & 5532.985 & -21923.165\cr
-424.63 & -308.244 & -386.653 & 125.67 & -74.709 & 273.379 & 250.767 & 48.032 & -378.848 & 53.352\cr
-824.755 & 638.712 & -1287.206 & 628.491 & 250.047 & 4601.007 & -3463.562 & 142.212 & 6594.3 & -3256.938\cr
-7934.871 & 1405.406 & -6188.958 & 2288.342 & -1669.7 & 3594.294 & -15306.448 & 813.979 & -10852.434 & 24785.914}
}.
\end{equation}
The weights obtained for $G_{Ep}/G_D$:
\begin{equation}
\vec{w}^T_{MP}=\scriptsize{(3.930227, 0.1108384, -5.325479, -2.846154, -0.2071328, 0.8742101, 0.4283194, 2.568322, 2.577635, -1.185632)}
\end{equation}
with the covariance matrix:
\begin{equation}
A^{-1} = \tiny{\pmatrix{
36866.41 & -52184.005 & 17354.195 & -9375.943 & 693.571 & 7949.39 & -16875.298 & -11986.299 & 10541.393 & 4687.308\cr
-68432.176 & 103227.83 & -25251.786 & 22514.051 & -1329.157 & -14150.394 & 34458.11 & 24954.878 & -19958.73 & -11910.498\cr
14227.233 & -16215.276 & 26518.973 & 2749.674 & 160.818 & 2066.422 & -2921.857 & 2228.451 & 2430.186 & -38.278\cr
-35928.337 & 57408.998 & -6198.14 & 18394.036 & -745.922 & -7749.442 & 20171.669 & 8522.062 & -11194.428 & -7642.097\cr
181.767 & -354.001 & 27.284 & -99.739 & 55.716 & -65.584 & -118 & -98.718 & 696.136 & -337.281\cr
6881.763 & -8970.135 & 2549.152 & -1714.537 & 128.027 & 2720.834 & -3012.463 & -2059.728 & 2495.438 & -329.115\cr
-23847.659 & 36615.2 & -6410.826 & 8820.603 & -475.037 & -5101.518 & 12518.641 & 9338.177 & -7159.334 & -4418.53\cr
22789.127 & -35227.846 & 11996.228 & -14431.928 & 441.824 & 4699.586 & -11630.739 & 8794.696 & 6644.126 & 4132.854\cr
3700.817 & -6332.644 & 791.595 & -1652.194 & 760.047 & -46.633 & -2130.248 & -1683.296 & 9852.346 & -4813.821\cr
17126.803 & -26781.151 & 4320.733 & -6631.983 & -137.16 & 3552.551 & -9216.18 & -6918.134 & -1252.964 & 7966.577}
}
\end{equation}
The weights obtained for $G_{Mp}/\mu_p G_D$:
\begin{equation}
\vec{w}^T_{MP}= \scriptsize{(-2.862682, -1.560675, 2.321148, 0.1283189, -0.2803566, 2.794296, 1.726774, 0.861083, 0.4184286, -0.1526676)}
\end{equation}
with the covariant matrix:
\begin{equation}
A^{-1} = \tiny{\pmatrix{
15709.171 & 6861.227 & 2766.185 & -6126.712 & -121.495 & 1318.866 & 8737.945 & 4008.92 & -94.694 & -3978.556\cr
3284.282 & 2803.079 & 843.705 & -1333.839 & -40.306 & 438.59 & 817.679 & 1156.474 & -31.955 & -1145.984\cr
1993.96 & 1142.807 & 495.778 & -954.548 & -31.697 & 341.671 & 836.303 & 462.66 & -25.013 & -454.17\cr
-3841.3 & -1694.722 & -859.519 & 2450.449 & 47.229 & -515.322 & -1122.054 & -167.405 & 37.338 & 155.266\cr
-88.252 & -54.872 & -31.981 & 50.267 & 8.865 & -76.635 & -33.677 & -16.192 & 6.739 & 12.676\cr
935.495 & 585.08 & 339.546 & -538.917 & -76.046 & 720.701 & 349.457 & 169.075 & -53.896 & -145.686\cr
14256.242 & 5279.888 & 2235.735 & -4498.246 & -93.725 & 1012.932 & 9909.287 & 4366.293 & -72.416 & -4343.104\cr
5227.654 & 2640.608 & 867.323 & -1228.301 & -32.179 & 348.563 & 3580.716 & 2148.384 & -24.917 & -2140.484\cr
-67.875 & -43.08 & -25.232 & 39.496 & 6.797 & -54.859 & -24.96 & -12.175 & 6.271 & 8.358\cr
-5204.745 & -2625.836 & -858.408 & 1214.946 & 28.58 & -324.213 & -3572.214 & -2144.257 & 21.06 & 2139.095}
}.
\end{equation}


\begin{thebibliography}{99}


\bibitem{Close_EM_book} F. E. Close, A. Donnachie, and G. Shaw, \textit{Electromagnetic Interactions and Hadronic Structure (Cambridge Monographs on Particle Physics, Nuclear Physics and Cosmology)},\\
 Cambridge 2007.


\bibitem{Lomon_VMD}
  G.~Hohler, E.~Pietarinen, I.~Sabba Stefanescu, F.~Borkowski, G.~G.~Simon, V.~H.~Walther and R.~D.~Wendling,
  \textit{Analysis Of Electromagnetic Nucleon Form-Factors},
  Nucl.\ Phys.\  B {\bf 114} (1976) 505; E. L. Lomon, Phys. Rev. {\bf C64} (2001) 035204; ibid {\bf C66} (2002) 045501;
 C.~Crawford {\it et al.},
  \textit{The Role of Mesons in the Electromagnetic Form Factors of the Nucleon},
  arXiv:1003.0903 [nucl-th].

\bibitem{Belushkin:2006qa}
  M.~A.~Belushkin, H.~W.~Hammer and U.~G.~Meissner,
  \textit{Dispersion analysis of the nucleon form factors including meson continua},
  Phys.\ Rev.\  C {\bf 75} (2007) 035202.




\bibitem{quark_model}
  G.~A.~Miller,
  \textit{Light front cloudy bag model: Nucleon electromagnetic form factors},
  Phys.\ Rev.\  C {\bf 66} (2002) 032201;
  F.~Cardarelli and S.~Simula,
  \textit{SU(6) breaking effects in the nucleon elastic electromagnetic form
  factors},
  Phys.\ Rev.\  C {\bf 62} (2000) 065201;
  R.~F.~Wagenbrunn, S.~Boffi, W.~Klink, W.~Plessas and M.~Radici,
  \textit{Covariant nucleon electromagnetic form factors from the  Goldstone-boson
  exchange quark model},
  Phys.\ Lett.\  B {\bf 511} (2001) 33;
  M.~M.~Giannini, E.~Santopinto and A.~Vassallo,
  \textit{An overview of the hypercentral constituent quark model}'
  Prog.\ Part.\ Nucl.\ Phys.\  {\bf 50} (2003) 263.

\bibitem{Perdrisat:2006hj}
  C.~F.~Perdrisat, V.~Punjabi and M.~Vanderhaeghen,
  \textit{Nucleon electromagnetic form factors},
  Prog.\ Part.\ Nucl.\ Phys.\  {\bf 59} (2007) 694.


\bibitem{Miller:2010nz}
  G.~A.~Miller, \textit{Transverse Charge Densities},
  arXiv:1002.0355 [nucl-th].



\bibitem{AlvarezRuso:2009mn}
  L.~Alvarez-Ruso,
  \textit{Theoretical highlights of neutrino-nucleus interactions},
  Plenary talk at 11th International Workshop on Neutrino Factories, Superbeams and Betabeams: NuFact09, Chicago, Illinois, 20-25 Jul 2009,
  arXiv:0911.4112 [nucl-th].


\bibitem{Alberico:2001sd}
  W.~M.~Alberico, S.~M.~Bilenky and C.~Maieron,
  \textit{Strangeness in the nucleon: Neutrino nucleon and polarized electron
  nucleon scattering},
  Phys.\ Rept.\  {\bf 358} (2002) 227.

\bibitem{Beck:2001dz}
  D.~H.~Beck and B.~R.~Holstein,
  \textit{Nucleon structure and parity-violating electron scattering},
  Int.\ J.\ Mod.\ Phys.\  E {\bf 10} (2001) 1.


\bibitem{Ahn:2006zza}
  M.~H.~Ahn {\it et al.}  [K2K Collaboration],
  \textit{Measurement of Neutrino Oscillation by the K2K Experiment},
  Phys.\ Rev.\  D {\bf 74} (2006) 072003
  [arXiv:hep-ex/0606032].


\bibitem{T2KLOI}
Y. Hayato et al., \textit{ Neutrino Oscillation Experiment at JHF, Letter of Intent to the JPARC
50 GeV Proton Synchrotron} (Jan. 21, 2003), {\tt http://neutrino.kek.jp/jhfnu/loi/loi JHFcor.pdf}

\bibitem{Llewellyn Smith:1971zm}
  C.~H.~Llewellyn Smith,
  \textit{Neutrino Reactions At Accelerator Energies}, Phys.\ Rept.\  {\bf 3} (1972) 261.


\bibitem{Gran:2006jn}
  R.~Gran {\it et al.}  [K2K Collaboration],
  \textit{Measurement of the quasi-elastic axial vector mass in neutrino oxygen interactions},
  Phys.\ Rev.\  D {\bf 74} (2006) 052002.



\bibitem{:2007ru}
  A.~A.~Aguilar-Arevalo {\it et al.}  [MiniBooNE Collaboration],
  \textit{Measurement of muon neutrino quasi-elastic scattering on carbon},
  Phys.\ Rev.\ Lett.\  {\bf 100} (2008) 032301.

\bibitem{Bernard:2001rs}
  V.~Bernard, L.~Elouadrhiri and U.~G.~Meissner,
  \textit{Axial structure of the nucleon},
  J.\ Phys.\ G {\bf 28} (2002) R1.

\bibitem{Kuzmin:2007kr}
  K.~S.~Kuzmin, V.~V.~Lyubushkin and V.~A.~Naumov,
  \textit{Quasielastic axial-vector mass from experiments on neutrino-nucleus scattering},
  Eur.\ Phys.\ J.\  C {\bf 54} (2008) 517.

\bibitem{Bodek:2007vi}
  A.~Bodek, S.~Avvakumov, R.~Bradford and H.~Budd,
  \textit{Extraction of the Axial Nucleon Form Factor from Neutrino Experiments on Deuterium},
  J.\ Phys.\ Conf.\ Ser.\  {\bf 110} (2008) 082004.



\bibitem{Alberico:1997vh}
  W.~M.~Alberico {\it et al.},
  \textit{Inelastic nu and anti-nu scattering on nuclei and *strangeness* of the nucleon},
  Nucl.\ Phys.\  A {\bf 623} (1997) 471.


\bibitem{Kim:2008zzc}
  K.~S.~Kim, M.~K.~Cheoun and B.~G.~Yu,
  \textit{Effect of strangeness for neutrino (anti-neutrino) scattering in the quasi-elastic region},
  Phys.\ Rev.\  C {\bf 77} (2008) 054604.

\bibitem{strangness}
  J.~Liu, R.~D.~McKeown and M.~J.~Ramsey-Musolf,
  \textit{Global Analysis of Nucleon Strange Form Factors at Low $Q^2$},
  Phys.\ Rev.\  C {\bf 76} (2007) 025202;
  R.~D.~Young, J.~Roche, R.~D.~Carlini and A.~W.~Thomas,
  \textit{Extracting nucleon strange and anapole form factors from world data},
  Phys.\ Rev.\ Lett.\  {\bf 97} (2006) 102002.




\bibitem{Bosted:1994tm}
  P.~E.~Bosted,
  \textit{An Empirical fit to the nucleon electromagnetic form-factors},
  Phys.\ Rev.\  C {\bf 51} (1995) 409;

\bibitem{Brash:2001qq}
  E.~J.~Brash, A.~Kozlov, S.~Li and G.~M.~Huber,
  \textit{New empirical fits to the proton electromagnetic form factors},
  Phys.\ Rev.\  C {\bf 65} (2002) 051001.

\bibitem{Budd:2003wb}
  H.~Budd, A.~Bodek and J.~Arrington,
  \textit{Modeling quasi-elastic form factors for electron and neutrino scattering}, Presented at 2nd International Workshop on Neutrino - Nucleus Interactions in the Few GeV Region (NUINT 02), Irvine, California, 12-15 Dec 2002.
  arXiv:hep-ex/0308005.

\bibitem{Arrington:2003df}
  J.~Arrington,
  \textit{How well do we know the electromagnetic form factors of the proton?},
  Phys.\ Rev.\  C {\bf 68} (2003) 034325.


\bibitem{Kelly:2004hm}
  J.~J.~Kelly,
  \textit{Simple parametrization of nucleon form factors},
  Phys.\ Rev.\  C {\bf 70} (2004) 068202.


\bibitem{Arrington:2006hm}
  J.~Arrington and I.~Sick,
  \textit{Precise determination of low-Q nucleon electromagnetic form factors and
  their impact on parity-violating e p elastic scattering},
  Phys.\ Rev.\  C {\bf 76} (2007) 035201.


\bibitem{Bodek:2007ym}
  A.~Bodek, S.~Avvakumov, R.~Bradford and H.~Budd,
  \textit{Vector and Axial Nucleon Form Factors:A Duality Constrained parameterization},
  Eur.\ Phys.\ J.\  C {\bf 53} (2008) 349.



\bibitem{Galster:1971kv}
  S.~Galster, H.~Klein, J.~Moritz, K.~H.~Schmidt, D.~Wegener and J.~Bleckwenn,
  \textit{Elastic electron - deuteron scattering and the electric neutron form-factor at four momentum transfers 5-fm**-2 < q**2 < 14-fm**-2},
  Nucl.\ Phys.\  B {\bf 32} (1971) 221.


\bibitem{Krutov:2002tp}
  A.~F.~Krutov and V.~E.~Troitsky,
  \textit{Extraction of the neutron charge form factor from the charge  form factor of deuteron},
  Eur.\ Phys.\ J.\  A {\bf 16} (2003) 285.


\bibitem{Alberico:2008sz}
  W.~M.~Alberico, S.~M.~Bilenky, C.~Giunti and K.~M.~Graczyk,
  \textit{Electromagnetic form factors of the nucleon: new fit and analysis of uncertainties},
  Phys.\ Rev.\  C {\bf 79} (2009) 065204.


\bibitem{Bishop_book} C. M. Bishop, \textit{Neural Networks for Pattern Recognition}, Oxford University Press 2008.

\bibitem{Geman92} S. Geman, E. Bienenstock, and R.  Doursat,
\textit{Neural networks  and the bias/variance dilema},
Neural Computation \textbf{4} (1), 1 (1992).






\bibitem{NN_hep_old} B. Denby,
\textit{Neural networks and cellular automata in experimental high energy physics},
Computer Physics Communications 49 (1988), 429;

\bibitem{NN_classify} Mellado B. et al.,
\textit{Prospects for the observation of a Higgs boson with $H\to\tau^+\tau^-\to l^+l^-\not{p_t}$ associated with one jet at the LHC},
Phys. Lett. B611 (2005),  60.


\bibitem{NN_compass} K. Kurek, E. Rondio, R. Sulej, K. Zaremba,
\textit{Application of the neural networks in events classification in the measurement of spin structure of the deuteron},
Meas. Sci. Technol. 18 (2007)  2486.

\bibitem{Forte:2002fg}
  S.~Forte, L.~Garrido, J.~I.~Latorre and A.~Piccione,
  \textit{Neural network parametrization of deep-inelastic structure functions},
  JHEP {\bf 0205} (2002) 062.


\bibitem{NN_esimator} J. Damgov and L. Litov,
\textit{Application of Neural Networks for Energy Reconstruction},
Nucl. Inst. Meth. A482 (2002) 776.





\bibitem{NNPDF} NNPDF Collaboration, \href{http://sophia.ecm.ub.es/nnpdf/}{http://sophia.ecm.ub.es/nnpdf/}.

\bibitem{Del Debbio:2004qj}
  L.~Del Debbio, S.~Forte, J.~I.~Latorre, A.~Piccione and J.~Rojo  [NNPDF
                  Collaboration],
  \textit{Unbiased determination of the proton structure function F2(p) with
  faithful uncertainty estimation},
  JHEP {\bf 0503} (2005) 080.

\bibitem{DelDebbio:2007ee}
  L.~Del Debbio, S.~Forte, J.~I.~Latorre, A.~Piccione and J.~Rojo  [NNPDF
                  Collaboration],
  \textit{Neural network determination of parton distributions: the nonsinglet
  case},
  JHEP {\bf 0703} (2007) 039.


\bibitem{Ball:2008by}
  R.~D.~Ball {\it et al.}  [NNPDF Collaboration],
  \textit{A determination of parton distributions with faithful uncertainty estimation},
  Nucl.\ Phys.\  B {\bf 809} (2009) 1;
  [Erratum-ibid.\  B {\bf 816} (2009) 293].


\bibitem{Ball:2009qv}
  R.~D.~Ball, L.~Del Debbio, S.~Forte, A.~Guffanti, J.~I.~Latorre, J.~Rojo and M.~Ubiali
                  [NNPDF Collaboration],
  \textit{Fitting Experimental Data with Multiplicative Normalization Uncertainties},
  arXiv:0912.2276 [hep-ph].


\bibitem{Ball:2010de}
  R.~D.~Ball, L.~Del Debbio, S.~Forte, A.~Guffanti, J.~I.~Latorre, J.~Rojo and M.~Ubiali,
  \textit{A first unbiased global NLO determination of parton distributions and their
  uncertainties},
  arXiv:1002.4407 [hep-ph].


\bibitem{MacKay92a}D. J. C. MacKay, \textit{Bayesian interpolation},
Neural Computation 4 (3), (1992) 415.

\bibitem{MacKay92b}D. J. C. MacKay, \textit{A practical Bayesian framework for backpropagation networks},
Neural Computation 4 (3), (1992) 448.





\bibitem{MacKay94}D. J. C. MacKay,
\textit{Bayesian methods for backpropagation networks},
in E. Domany, J. L. van Hemmen, and K. Schulten (Eds.),
\textit{Models of Neural Networks III}, Sec. 6. New York: Springer-Verlag (1994).


\bibitem{sulej_network} \textit{NetMaker} \href{http://www.ire.pw.edu.pl/~rsulej/NetMaker/}{http://www.ire.pw.edu.pl/$\sim$rsulej/NetMaker/}
(written in C\#); raw fit results presented in this paper available at
\href{http://www.ire.pw.edu.pl/~rsulej/NetMaker/index.php?pg=h33}{http://www.ire.pw.edu.pl/$\sim$rsulej/NetMaker/index.php?pg=h33}.


\bibitem{Kolmogorov} V. Kurková, \textit{Kolmogorov's theorem and multilayer neural networks},
Neural Networks 5, Nr 3 (1992), 501.
\bibitem{NN_weight_decay} A. S. Weigend, D. E. Rumelhart, B. A. Huberman,
\textit{Generalization by Weight-Elimination with Application to Forecasting},
Proceedings of the Conference on Advances in Neural Information Processing Systems, Vol. 3 (1990), pp. 875-882, Denver, Colorado, US.


\bibitem{NN_approx1} K. Hornik, \textit{Approximation Capabilities of Multilayer Feedforward Networks},
Neural Networks 4, Nr 2 (1991), 251.

\bibitem{NN_approx2} M. Leshno et al.,
\textit{Multilayer Feedforward Networks With a Nonpolynomial Activation Function Can Approximate Any Function},
Neural Networks 6, Nr 6 (1993), 861.


\bibitem{mlp} D. E. Rumelhart et al, \textit{Learning internal representations by error propagation},
monograph D. E. Rumelhart and J. A. McClelland Parallel Distribuited Processing: \textit{Exploration in the Microstructure of Cognition},
Vol. 1 (1986), pp 318-362, The MIT Press.




\bibitem{NN_Levenberg} K. Levenberg, \textit{A Method for the Solution of Certain Non-Linear Problems in Least Squares},
The Quarterly of Applied Mathematics 2 (1944), 164.

\bibitem{NN_Marquardt} D. Marquardt, \textit{An Algorithm for Least-Squares Estimation of Nonlinear Parameters},
SIAM Journal on Applied Mathematics 11 (1963), 431.

\bibitem{NN_quickprop} S. Fahlman, \textit{An Empirical Study of Learning Speed in Back-Propagation Networks},
CMU-CS-88-162, School of Computer Science, Carnegie Mellon University, (1988).

\bibitem{RPROP} Ch. Igel, \textit{Improving the Rprop Learning Algorithm} Proceedings of the Second International Symposium on Neural Computation, NC'2000, pp. 115-121, ICSC Academic Press, 2000.


\bibitem{Thodberg}
H. H. Thodberg, \textit{Ace of Bayes: application of neural networks with pruning},
Technical Report 1132E, The Danish Meat Research Institute, Maglegaardsvej 2, DK-4000 Roskilde, Denmark. 1993.


\bibitem{Neal94} R. M. Neal,
\textit{Bayesian Learning for Neural Networks}. Ph.D thesis, University of Toronto, Canada.


\bibitem{bishop_hessian} C.M. Bishop,
\textit{Exact calculation of the Hessian matrix for the multilayer perceptron},
Neural Computation 4 (4), 494 (1992).

\bibitem{Gull88b} S. F. Gull, \textit{Bayesian inductive inference and maximum entropy.} In G. J. Ericson and C. R. Smith (Eds.)
Maximum-Entropy and Bayesian Methods in Science and Engineering, Vol. 1: Fundations, pp 53-74 (1988). Dordrecht: Kluwer.
S. F. Gull, \textit{Development in maximum entropy data analysis}. In J. Skilling (Ed.), \textit{Maximum Entropy and Bayesian Methods},
Cambridge, 1988, pp. 53-71. Dordrecht: Kluwer.

\bibitem{Williams} P. M. Williams, \textit{Bayesian Regularization and Pruning using a Laplace Prior}, Neural Computation \textbf{7} (1), 117 (1995).




\bibitem{Arrington:2007ux}
  J.~Arrington, W.~Melnitchouk and J.~A.~Tjon,
  \textit{Global analysis of proton elastic form factor data with two-photon exchange
  corrections},
  Phys.\ Rev.\  C {\bf 76} (2007) 035205.

\bibitem{Qattan:2004ht}
  I.~A.~Qattan {\it et al.},
  \textit{Precision Rosenbluth measurement of the proton elastic form factors},
  Phys.\ Rev.\ Lett.\  {\bf 94} (2005) 142301.

\bibitem{Gayou:2001qt}
  O.~Gayou {\it et al.},
  \textit{Measurements of the elastic electromagnetic form factor ratio mu pgep/gmp
  via polarization transfer},
  Phys.\ Rev.\  C {\bf 64} (2001) 038202.




\bibitem{TPE_papers}
  P.~A.~M.~Guichon and M.~Vanderhaeghen,
  \textit{How to reconcile the Rosenbluth and the polarization transfer method in the measurement of the proton form factors},
  Phys.\ Rev.\ Lett.\  {\bf 91} (2003) 142303;
  P.~G.~Blunden, W.~Melnitchouk and J.~A.~Tjon,
  \textit{Two-photon exchange and elastic electron proton scattering},
  Phys.\ Rev.\ Lett.\  {\bf 91} (2003) 142304;
  Y.~C.~Chen, A.~Afanasev, S.~J.~Brodsky, C.~E.~Carlson and M.~Vanderhaeghen,
\textit{Partonic calculation of the two-photon exchange contribution to elastic electron proton scattering at large momentum transfer},
  Phys.\ Rev.\ Lett.\  {\bf 93} (2004) 122301;
  A.~V.~Afanasev, S.~J.~Brodsky, C.~E.~Carlson, Y.~C.~Chen and M.~Vanderhaeghen,
  \textit{The two-photon exchange contribution to elastic electron nucleon scattering at large momentum transfer},
  Phys.\ Rev.\  D {\bf 72} (2005) 013008.



\bibitem{Carlson:2007sp}
  C.~E.~Carlson and M.~Vanderhaeghen,
  \textit{Two-photon physics in hadronic processes},
  Ann.\ Rev.\ Nucl.\ Part.\ Sci.\  {\bf 57} (2007) 171.






\bibitem{proton_form_factor_data} L. Andivahis et al., Phys. Rev. D 50, 5491 (1994);
W. Bartel et al., Nuclear Physics B58, 429,(1973);
Ch. Berger et al., Phys Letters 35B, 87-89, (1971);
F.Borkowski et al., Nucl Phys B93, 461-478,(1975);
K.M. Hanson, et al., Phys Rev D, vol. 8, no. 3, 753-778,(1973);
L.E. Price, et al., Phys Rev D4, 45-53,(1971);
R.C. Walker et al., Phys Rev. D 49, 5671 (1994).


\bibitem{the_rest_of_GMp_olddata} P.E.Bosted,et al.,Phys Rev C42,38-64,(1990)
A.~F. Sill et al., PRD 48, 29-55(1993).

\bibitem{the_rest_of_GEp_olddata} G.G. Simon et al., Nucl. Phys. A, 381-391 (1979);
J.J. Murphy et al., Phys Rev C9, 2125-2129 (1974).

\bibitem{JLab_data} \href{http://www.jlab.org/resdata}{http://www.jlab.org/resdata}.



\bibitem{BLAST:2008ha}
  E.~Geis {\it et al.}  [BLAST Collaboration],
  \textit{The Charge Form Factor of the Neutron at Low Momentum Transfer from the
  $^{2}\vec{\rm H}(\vec{\rm e},{\rm e}'{\rm n}){\rm p}$ Reaction},
  Phys.\ Rev.\ Lett.\  {\bf 101} (2008) 042501.

\bibitem{Xu:2002xc}
  W.~Xu {\it et al.}  [Jefferson Lab E95-001 Collaboration],
  \textit{PWIA extraction of the neutron magnetic form factor from quasi-elastic He-3(pol.)(e(pol.),e') at Q**2 = 0.3-(GeV/c)**2 to 0.6-(GeV/c)**2},
  Phys.\ Rev.\  C {\bf 67} (2003) 012201.





\bibitem{granet} \href{http://www.ift.uni.wroc.pl/~kgraczyk/nn.html}{http://www.ift.uni.wroc.pl/$\sim$kgraczyk/nn.html}.













\end{thebibliography}
\end{document}